\documentclass[10pt, twocolumn]{extarticle}
\usepackage{graphicx}
\usepackage{hyperref}
\usepackage{amssymb}
\usepackage{amsmath}
\usepackage{amsfonts}
\usepackage{parskip}
\usepackage{titling}
\usepackage[table]{xcolor}
\usepackage[square,sort&compress,comma,numbers]{natbib}
\usepackage{multirow}
\usepackage{empheq}

\usepackage{titlesec}
\usepackage{caption}
\titleformat{\section}{\large\bf}{\thesection}{1em}{}
\titleformat{\subsection}{\bf}{\thesubsection}{1em}{}
\titleformat{\subsubsection}{\it}{\thesubsubsection}{1em}{}

\pretitle{\Large\bf\vskip 0.3em} \posttitle{\par\vskip 0.6em} \preauthor{\large} \postauthor{\par\vskip -0.0em}
\predate{\large} \postdate{\par\vskip 0.1em \rule{\linewidth}{0.5pt}} 

\newcommand{\beq}{\begin{equation}}
\newcommand{\eeq}{\end{equation}}
\newcommand{\bea}{\begin{eqnarray}}
\newcommand{\eea}{\end{eqnarray}}

\newcommand{\comment}[1]{}
\renewcommand{\d}{{\rm d}}

\newcommand{\sinc}{{\rm sinc}}

\begin{document}
\captionsetup{font=small}

\title{Hybrid AM/FM Mode-Locking of Singly-Resonant OPOs}
\author{Ryan Hamerly$^{1,2}$, Evan Laksono$^3$, Marc Jankowski$^{2,3}$, Edwin Ng$^{2,3}$, Noah Flemens$^3$, Myoung-Gyun Suh$^{2}$, and Hideo Mabuchi$^3$}
\date{\today}


\maketitle



\begin{flushleft}
\small
$^{1}$      \textit{Research Laboratory of Electronics, MIT, 50 Vassar Street, Cambridge, MA 02139, USA} \\
$^{2}$      \textit{NTT Research Inc., Physics and Informatics Laboratories, 940 Stewart Drive, Sunnyvale, CA 94085, USA} \\
$^{3}$      \textit{E.~L.~Ginzton Lab, Stanford University, 348 Via Pueblo Mall, Stanford, CA 94305, USA}
\end{flushleft}

{\bf\small 
Abstract---We investigate a new mode-locking regime in the singly-resonant OPO employing simultaneous amplitude- and frequency-modulation of the intracavity field.  This OPO exhibits deterministic, ``turn-key'' formation of a stable, broadband, chirped frequency comb with high conversion efficiency.  Comb-forming dynamics follow a simple phase-space dynamical model, governed by cavity dispersion and modulator chirp, which agrees closely with full numerical simulations.  The comb exhibits fast, mode-hop-free tuning over the full gain window of the OPA crystal, controlled by the modulator frequency.  Conditions for comb stability, and techniques to enhance comb bandwidth through intentional phase-mismatch and chirping, are investigated.  
}

\rule{\linewidth}{0.5pt}

\section{Introduction}

The optical frequency comb is a cornerstone of precision science \cite{diddams2010evolving}, with applications in metrology \cite{poli2013optical}, communications \cite{pfeifle2014coherent, rizzo2022petabit}, spectroscopy \cite{coddington2016dual}, and laser ranging \cite{baumann2014comb, trocha2018ultrafast}.  Recent advances in photonic fabrication, together with the desire for compact, low-cost, mass-produced comb sources, has made miniaturized, on-chip ``micro-combs'' an imminent possibility \cite{pasquazi2018micro, chang2022integrated}---with the most prominent approaches based on either Kerr nonlinearity \cite{kippenberg2004kerr, savchenkov2004low}, electro-optic modulation (EOM) \cite{torres2014optical, kobayashi1972high, ho1993optical, zhang2019broadband}, or mode-locking of lasers (MLL) \cite{haus2000mode, hillbrand2020phase}.  The main approaches to micro-combs all suffer from various limitations---for example, MLL combs have limited free-spectral range (FSR) and gain-limited bandwidth \cite{chang2022integrated}, while Kerr and EOM combs usually suffer from low efficiency \cite{coen2013universal} (but see \cite{kim2019turn, hu2022high}), require high cavity $Q \sim 10^6$, and involve precise locking of the pump to a cavity resonance.  Kerr combs also exhibit complex, chaotic comb-forming dynamics \cite{shen2020integrated}, while EOM combs may have simpler dynamics but require high radio-frequency (RF) power consumption \cite{torres2014optical}.  Because of this, there is no ideal comb source, and micro-comb users must navigate the harsh tradeoff between the benefits and drawbacks of each platform.

These challenges has fueled continued interest in combs based on parametric ($\chi^{(2)}$) nonlinearities \cite{bruch2021pockels, englebert2021parametrically}, specifically in the context of optical parametric oscillators (OPOs) \cite{boyd2008nonlinear}.  The inherently broadband $\chi^{(2)}$ gain in OPOs, combined with chirped poling \cite{suchowski2014adiabatic} and/or dispersion engineering \cite{jankowski2021dispersion}, allows for efficient, versatile, broadly tunable light generation \cite{harris1969tunable, eckardt1991optical}, including at wavelengths where we lack good laser gain media.  Synchronous pumping is the most common method to generate an OPO comb \cite{van1995synchronously, cheung1990theory, hamerly2016reduced, jankowski2018temporal, roy2022temporal}, but requires that the pump itself be a comb source.  Diddams {\it et~al.}\ demonstrated comb generation from a continuous-wave (CW)-pumped OPO \cite{diddams1999broadband} by intracavity frequency-modulated (FM) mode-locking \cite{harris1964fm}, a technique that has attracted follow-on work and recent on-chip implementations \cite{forget2006actively, melkonian2007active, esteban2012frequency, stokowski2024integrated}.  However, the doubly-resonant nature of this device is a major limitation: to date the FM-OPO has only been demonstrated near degeneracy, with limited tuning range.  Moreover, the pump must be accurately locked to the second-harmonic of the cavity resonance, making the OPO very sensitive to wavelength drift.  Despite its promise, FM mode-locking alone does not unlock the full range of capabilities one would expect from an OPO comb.

In this article, we propose a new comb-generation mechanism for {\it singly-resonant} OPOs based on simultaneous amplitude-modulated (AM) and FM mode locking.  This hybrid AM/FM mode-locked OPO, which we call the {\it quadrature-amplitude modulated OPO} (QAM-OPO), shares the main advantages of the FM-OPO: broad comb bandwidth, high efficiency, CW (as opposed to frequency comb) pump, and deterministic ``turn-key'' operation \cite{stokowski2024integrated}.  However, the singly-resonant condition provides additional benefits not present in the FM-OPO: it has a much flatter power spectrum, it can work away from degeneracy without sacrificing comb bandwidth, it is broadly tunable, and it is insensitive to pump phase noise or frequency drift, with pump phase fluctuations imparted onto the non-resonant idler.  Moreover, the presence of an idler further extends the bandwidth of the comb.  Taken together, these attributes make the QAM-OPO a promising source for stable, broadband, widely tunable frequency combs, fully harnessing the strengths of the $\chi^{(2)}$ nonlinearity.

This paper is structured as follows.  Sec.~\ref{sec:model} presents the basic theory of the QAM-OPO: starting with the nonlinear field equations and invoking a frequency-modulated continuous-wave (FMCW) ansatz, we obtain a reduced phase-space dynamical model that accurately predicts the dynamics and steady-state behavior when the OPO is in a stable operating regime.  In general, this model predicts a highly-chirped, $O(1)$ duty-cycle pulse train with a very flat optical power spectrum, and by analogy to the inverted pendulum, we derive an analytic expression for the comb bandwidth (as an illustrative example, we simulate FMCW mode-locking in a bulk cavity, comprising discrete modulators and fiber feedback sections, although the models developed here are readily extended to integrated realizations, such as in thin-film lithium niobate \cite{zhu2021integrated}).  Sec.~\ref{sec:stability} analyzes the stability of this comb source and highlights three phenomena---wraparound, loopback, and mode-hopping---that can destabilize the comb.  This analysis provides analytic guideposts to ensure stable comb generation, as well as bounds on important properties including bandwidth, FSR, and number of comb lines.  Finally, Sec.~\ref{sec:tuning} explores the tunability of the QAM-OPO, where we show that the comb center frequency can be directly controlled through the EOM's RF detuning.  Moreover, this tuning range can be increased if we have control over the pump wavelength.  Sec.~\ref{sec:conc} states our conclusions.

Three Appendices cover supplementary topics that may be helpful to the reader.  Appendix~\ref{sec:not} covers the basic theory of OPOs and notation used in this paper.  Appendix.~\ref{sec:bw} covers techniques to maximize the $\chi^{(2)}$ phase-matching bandwidth.  Finally, Appendix~\ref{sec:deg} considers the degenerate or near-degenerate case (where an OPO can be made singly-resonant with a periodic filter).  

\section{Basic Model of QAM-OPO} \label{sec:model}

\subsection{EO Comb vs.\ OPO Comb}

At a high level, we can motivate the QAM-OPO by comparison to a resonant electro-optic (EO) comb.  A conventional EO comb consists of a ring cavity with an intracavity phase modulator (PM), driven at the cavity FSR (or a harmonic thereof), which induces coupling between neighboring cavity modes, and converting the CW pump into a frequency comb (Fig.~\ref{fig:f1}(a)).  Consider a cavity with round-trip loss $\alpha$, PM modulation frequency $\Omega$ and amplitude $\phi_p$ (i.e.\ modulation $e^{i \phi_p\cos(\Omega t)}$).  
Under a simple dispersionless coupled-mode model, the power of the $m^{\rm th}$ ($m \neq 0$) comb line is given by $P_m = P_{\rm in} (\alpha/\phi_p) e^{-|m|\alpha/\phi_p}$ \cite{zhang2019broadband, buscaino2020design}.  Both the 3-dB bandwidth $N_{\rm 3dB}$ (defined in terms of number of comb lines) and conversion efficiency $\eta_{\rm comb}$ both depend on the figure of merit $\phi/\alpha$:
\beq
	N_{\rm 3dB} = 2\log(2)\phi_p/\alpha,\ \ \ 
	\eta_{\rm comb} = 2\alpha/\phi_p \label{eq:2-neta}
\eeq
Eq.~(\ref{eq:2-neta}) highlights two limitations of EO combs.  (1) First, the comb generation efficiency is very poor and inversely related to bandwidth $\eta_{\rm comb} \propto N_{\rm 3dB}^{-1}$ (Kerr combs follow the same scaling law \cite{coen2013universal}).  This can be circumvented using coupled-cavity designs \cite{hu2022high, buscaino2020design}, but such systems pose unique locking challenges, particularly for materials like LiNbO$_3$ that suffer from photorefractive drift.  (2) Second, broad combs require very low cavity loss.  Rewriting in terms of drive voltage $V_{\rm pp}$ (see Appendix~\ref{sec:a-eo} for derivation), we find $N_{\rm 3dB} \approx V_{\rm pp}/(V_\pi L\alpha)$, 
i.e.\ a broad comb requires a very low modulator loss figure-of-merit $V_\pi L \alpha$, or a large $V_{\rm pp}$ requiring high RF power.

The OPO comb (Fig.~\ref{fig:f1}(b)) operates by a very different mechanism.  Here, in addition to the EOM, the cavity contains a $\chi^{(2)}$ gain element, which in the presence of a CW pump, facilitates gain, i.e.\ optical parametric amplification (OPA), via parametric down conversion (PDC) of pump photons to signal and idler photons.  As in the EO comb, the EOM induces coupling between cavity modes, producing the comb.  However, the OPO gain medium counteracts the cavity loss, allowing one to circumvent the $V_\pi L\alpha$-derived limits for conventional EO combs.  Moreover, the conversion efficiency of the OPO comb is derived from the CW OPO, is of order unity and not subject to an efficiency-bandwidth tradeoff.

\subsection{Ikeda Map}

\begin{figure}[b!]
\begin{center}
\includegraphics[width=1.00\columnwidth]{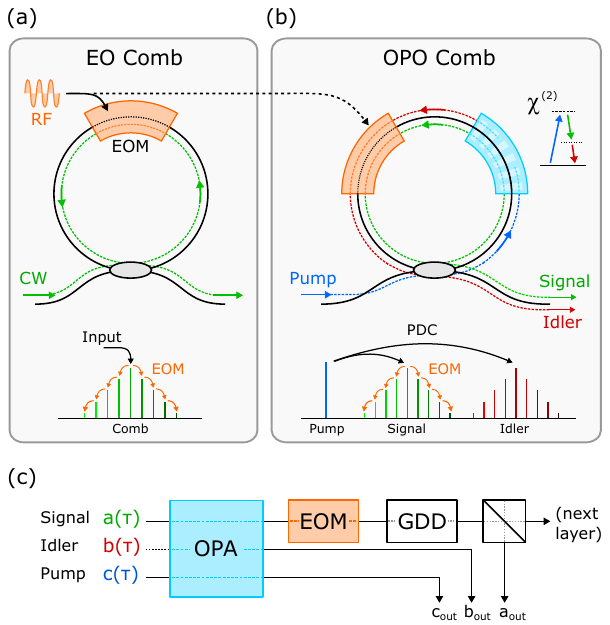}  
\caption{Comb Generation Mechanism. (a) Conventional resonant EO comb and (b) singly-resonant OPO comb. (c) Ikeda map round-trip model of OPO comb. 
}
\label{fig:f1}
\end{center}
\end{figure}

The field evolution of the OPO comb is governed by three forcing terms: (1) gain and loss, (2) modulator chirp, and (3) group-delay dispersion (GDD).  We model these dynamics using the Ikeda map, where the state of the cavity is encoded in the propagating signal field $a^{(n)}(\tau)$, a $T$-periodic function where $\tau$ is the ``fast'' time variable and $n$ is the round-trip number, and $T$ is the cavity repetition rate \cite{ikeda1979multiple}.  A round trip can be decomposed into a sequence of discrete operations as shown in Fig.~\ref{fig:f1}(c):
\begin{figure}[b!]
\begin{center}
\includegraphics[width=1.00\columnwidth]{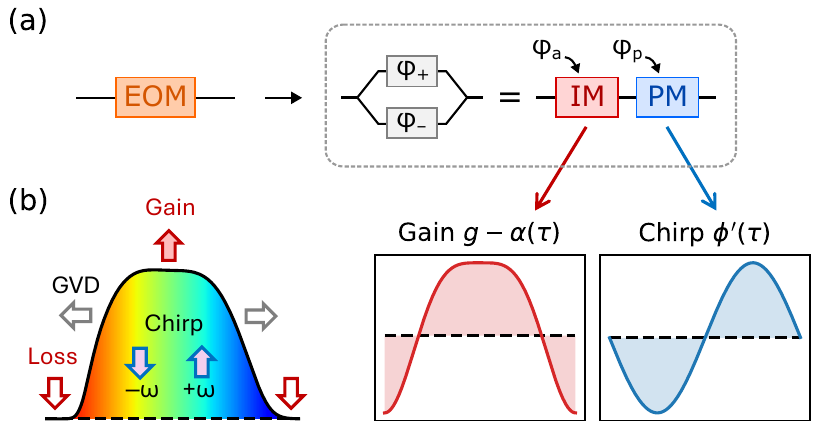}
\caption{(a) Simultaneous amplitude- and phase-modulation in QAM-OPO.  (b) Competition between gain, chirp, GVD, and loss, that stabilizes the comb output.}
\label{fig:f2}
\end{center}
\end{figure}

\begin{enumerate}
	\item {\it OPA.} In the $\chi^{(2)}$ crystal, the signal $a$, idler $b$, and pump $c$ (at carrier frequencies $\omega_a, \omega_b$, and $\omega_c = \omega_a + \omega_b$) mix in a three-wave interaction.  Dispersion must also be accounted for, and the equations are solved in the propagative picture \cite{boyd2008nonlinear}
\begin{subequations}\begin{align}
    \partial_z a & = i \beta_a(i\partial_\tau) a + i \kappa\, c b^* \label{eq:2-da} \\
    \partial_z b & = i \beta_b(i\partial_\tau) b + i \kappa\, c a^* \label{eq:2-db} \\
    \partial_z c & = \underbrace{i \beta_c(i\partial_\tau) c}_{\text{Dispersion}} + \underbrace{\vphantom{\beta_c(i\partial_\tau)} i \kappa\, a b}_{\text{OPA}} \label{eq:2-dc}
\end{align}\label{eq:2-d}\end{subequations}
over the domain $z \in [0, L]$, with initial conditions $b(\tau) = 0$, $c(\tau) = c_{\rm in}$.  Here $L$ is the OPA crystal length, $\kappa$ is the nonlinear coefficient, $c_{\rm in}(\tau)$ is the pump (a constant for CW pumping), and $\beta_{u}(\omega)$ ($u \in \{a, b, c\}$) are the dispersion relations for the three waves.  In general, Eqs.~(\ref{eq:2-d}) must be solved numerically.  A full treatment is given in Appendix~\ref{sec:not}, but for a CW signal field, the OPA crystal is modeled as a nonlinear gain element $a(\tau) \rightarrow e^{ig(a, \omega)/2} a(\tau)$, where the gain $g(a, \omega)$ is a function of the signal amplitude and its frequency.

In the high-finesse limit, $g(a, \omega)$ has an analytic expression (see Eq.~\ref{eq:a-gnorm}):
\beq
	g = p^2\alpha\, \text{sinc}^2\bigl(\sqrt{|\kappa a|^2 + (\Delta\beta/2)^2}\, L\bigr) \label{eq:2-gnorm}
\eeq
which depends on the pump normalized to threshold $p = c_{\rm in}/c_{\rm th}$, the round-trip loss $\alpha$, and the phase-mismatch $\Delta\beta$, which itself depends on $\omega \equiv \omega_a$:
\begin{align}
	\Delta\beta & = \beta_c - \beta_a - \beta_b \nonumber \\
		& = \beta(\omega_c) - \beta(\omega) - \beta(\omega_c-\omega) 
\end{align}
In the linear, phase-matched regime ($a = 0, \Delta\beta = 0$), the OPO is at threshold, i.e.\ $g = \alpha$.  Both phase mismatch and pump depletion add in quadrature to reduce the gain from its optimal value.

	\item {\it EOM.}  An OPO comb requires modulation---which in the QAM-OPO cascades an intensity-  and phase-modulator (IM and PM, Fig.~\ref{fig:f2}(a)):
	\beq
		a(\tau) \rightarrow \underbrace{\cos(\rho(\tau))}_{\text{IM}} \underbrace{\vphantom{\cos(\rho(\tau))}{}e^{i\phi(\tau)}}_{\text{PM}}
	\eeq
	where the EOM inputs $\rho(\tau), \phi(\tau)$ are sinusoidal fields with frequency $\Omega$ and amplitude $\phi_{a}, \phi_{p}$, with the IM bias set to maximize the transmission at the point of zero chirp ($\rho(0) = 0$):
	\beq
		\rho(\tau) = \bigl(\cos(\Omega\tau) - 1) \phi_a,\ \ 
		\phi(\tau) = (\cos(\Omega\tau)-1) \phi_p \label{eq:2-rp}
	\eeq
	One could also use a dual-drive Mach-Zehnder modulator (MZM), where $a(\tau) \rightarrow \tfrac12(e^{i\psi_+(\tau)}+e^{i\psi_-(\tau)}) a(\tau)$, to achieve QAM modulation.  Here, up to an overall phase constant, the inputs on the MZM arms are $\psi_\pm(\tau) = \rho(\tau) \pm \phi(\tau)$, corresponding to modulation amplitudes of $\phi_\pm = \phi_a \pm \phi_p$.  This modulator is shorter than the IM/PM cascade, but will require more phase and therefore higher drive voltage.
	\item {\it GDD.}  Dispersion is modeled with as a frequency-dependent phase.  The OPA and EOM are also dispersive, and in the high-finesse regime, we can treat all these dispersions as one lumped element, $a(\omega) \rightarrow e^{i\delta(\omega)} a(\omega)$, where $\delta(\omega)$ is the total  cavity round-trip dispersion.
	\item {\it Out-coupling.}  The field is attenuated  $a \rightarrow e^{-\alpha_0/2} a$.
\end{enumerate}

Taken together, these effects give us a qualitative model that explains comb formation in the QAM-OPO, sketched in Fig.~\ref{fig:f2}(b): (1) the IM creates a localized region of net gain, which seeds the comb, (2) the PM chirp red- and blue-shifts the tails of the pulse, and (3) dispersion (predominantly GDD) pushes power from these tails from the net gain region into the net-loss region.

\subsection{FMCW Ansatz and Phase-Space Model}
\label{sec:2-fmcw}
  
Here, we propose an ansatz describing the signal field in terms of a frequency-modulated continuous-wave (FMCW) function, i.e.\ a function $a(\tau)$ with amplitude $A(\tau)$ and carrier frequency $\omega(\tau)$ varying slowly in time:
\beq
	a^{(n)}(\tau) = A^{(n)}(\tau) \exp\Bigl[-i\Bigl(\phi_0 + \int_0^t{\omega^{(n)}(\tau') \d \tau'}\Bigr)\Bigr] \label{eq:2-fmcw}
\eeq
where we have explicitly denoted the round-trip index $n$, which will be important when deriving dynamical equations.  However, except when necessary, we drop this index in the results below.  

The FMCW ansatz is motivated by our empirical observation that, in CW-pumped OPO combs, the amplitude $|a(\tau)|$ is a slowly-varying quantity, making the waveform quasi-CW.  This is a unique property of these combs that stands in contrast with the bright solitons \cite{leo2010temporal, herr2014temporal}, non-soliton pulses \cite{zhang2019broadband}, dark solitons \cite{xue2015mode}, and simultons \cite{jankowski2018temporal} observed in conventional frequency-comb platforms, and arises because the comb is stabilized by balancing GDD with modulator {\it chirp}, rather than Kerr nonlinearity as in a conventional soliton comb.
  
\begin{table}[b!]
\begin{center}
\begin{tabular}{ll|c}
	\hline\hline
	\multicolumn{2}{c|}{\bf Property} & {\bf Default} \\ \hline
	OPA Length & $L$ & 1 cm \\
	EOM Length & $L_{\rm eo}$ & 1 cm \\
	OC Loss & $\alpha_0$ & 0.2 \\
	Pump / Threshold & $p$ & 1.65 \\
	IM, PM Phase & $\phi_a, \phi_p$ & -- \\
	EOM Frequency & $f_{\rm rep}$ & -- \\
	Disp.\ Comp.\ Factor & DCF & 1 \\
	\multicolumn{2}{l|}{OPA/EOM Material} & LiNbO$_3$ \\
	\multicolumn{2}{l|}{Disp.\ Comp.\ Material} & SMF28 fiber \\
	\hline\hline
\end{tabular}
\caption{Default settings for QAM-OPO used for simulations in this paper.}
\label{tab:t1}
\end{center}
\end{table}

\begin{table}[b!]
\begin{center}
\begin{tabular}{c|c|cl}
\hline\hline
 & {\bf C-Band} & {\bf O-Band} &  \\ \hline
$\lambda_a$        & 1.550 & 1.300 & \!\!\!\!\!\!$\mu$m \\ 
$\lambda_b$        & 2.362 & 2.802 & \!\!\!\!\!\!$\mu$m \\ 
$\lambda_c$        & 0.936 & 0.888 & \!\!\!\!\!\!$\mu$m \\ \hline
$n_{g,a}, n_{g,b}$ & 2.174 & 2.184 &  \\ 
$n_{g,c}$          & 2.222 & 2.231 &  \\ 
$\Delta\beta_{1}$    & 0.158 & 0.159 & \!\!\!\!\!\!ps/mm  \\ \hline
$\Lambda$          & 27.68 & 25.73 & \!\!\!\!\!\!$\mu$m \\ 
 \hline\hline
\end{tabular}
\caption{Group-velocity matched pumping conditions for bulk LiNbO$_3$ dispersion relation.  $\lambda_a/\lambda_b/\lambda_c$: signal/idler/pump.  $n_g = c\,\beta'(\omega)$: group index, $\Delta\beta_1 = \beta'(\omega_c)-\beta'(\omega_a)$: pump-signal GV mismatch, $\Lambda$: quasi phase-match (QPM) period.}
\label{tab:t2}
\end{center}
\end{table}

Using the Ikeda map, we wish to obtain a field equation for $A(n, \tau) \equiv A^{(n)}(\tau)$, $\omega(n, \tau) = \omega^{(n)}(\tau)$, where the round-trip index $n$ is converted into a continuous variable, and related to the ``slow'' time in the sense of Lugiato-Lefever by $t = n T$ \cite{lugiato1987spatial}.  This equation will be accurate when field is slowly-varying, so that we may replace $f^{(n+1)}-f^{(n)} \rightarrow \partial f/\partial n$, arriving at the following field equations for $A$ and $\omega$:
\begin{subequations}\begin{align}
	\frac{\partial A}{\partial n} & = \underbrace{-\delta'(\omega) \frac{\partial A}{\partial \tau}}_{\text{GDD}} + \underbrace{\frac{1}{2}\bigl[g(A, \omega) - \alpha(\tau)\bigr] A}_{\text{gain, loss}} \\
	\frac{\partial \omega}{\partial n} & = \underbrace{-\delta'(\omega) \frac{\partial\omega}{\partial \tau}}_{\text{GDD}} - \underbrace{\phi'(\tau) \vphantom{\frac\partial\partial}}_{\rm chirp} 
\end{align}\label{eq:2-daw}\end{subequations}
Here $\delta(\omega) = \beta(\omega)L + \delta_{\rm ext}(\omega)$ is the total GDD, which is a sum of OPA dispersion and other intracavity contributions, and $\phi(\tau)$ is the round-trip EOM phase shift.  The QAM-OPO also includes a round-trip intensity modulation, so the total loss (out-coupling plus modulation) $\alpha(\tau) = \alpha_0 + 2\log[\sec(\rho(\tau))]$ is time-dependent as well.

\begin{figure}[b!]
\begin{center}
\includegraphics[width=1.00\columnwidth]{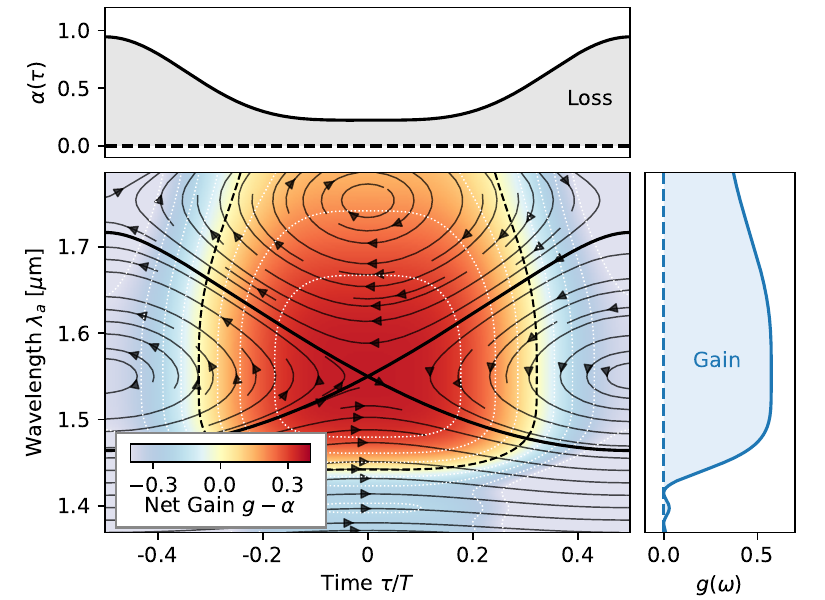}
\caption{Phase-space dynamics $\d\tau/\d n$, $\d\omega/\d n$ (streamlines) and net gain $g - \alpha$ (contours) for QAM-OPO.  Parameters: C-band GV-matched, DCF = 2, $\phi_a = 0.4$, $\phi_p = 1.6$.}
\label{fig:f3}
\end{center}
\end{figure}

Eqs.~(\ref{eq:2-daw}) can be reduced to a system of ordinary differential equations (ODEs) using the method of characteristics, where $(A, \omega)$ are integrated along characteristic curves $(n, \tau(n))$ as follows:
\begin{align}
	\frac{\d\tau}{\d n}   & =  \delta'(\omega),
	& \frac{\d\omega}{\d n} & = -\phi'(\tau),
	& \frac{\d A}{\d n}     & =  [g(A,\omega)-\alpha(\tau)]A \label{eq:2-char}
\end{align}
These equations provide quantitative foundation for the dynamical {\it phase-space} model of Fig.~\ref{fig:f2}(b), where energy flow is governed by gain/loss ($\d A/\d n$), group delay ($\d\tau/\d n$), EOM chirp ($\d\omega/\d n$), and the steady-state comb is the solution that balances these effects.

\begin{figure*}[t!]
\begin{center}
\includegraphics[width=1.00\textwidth]{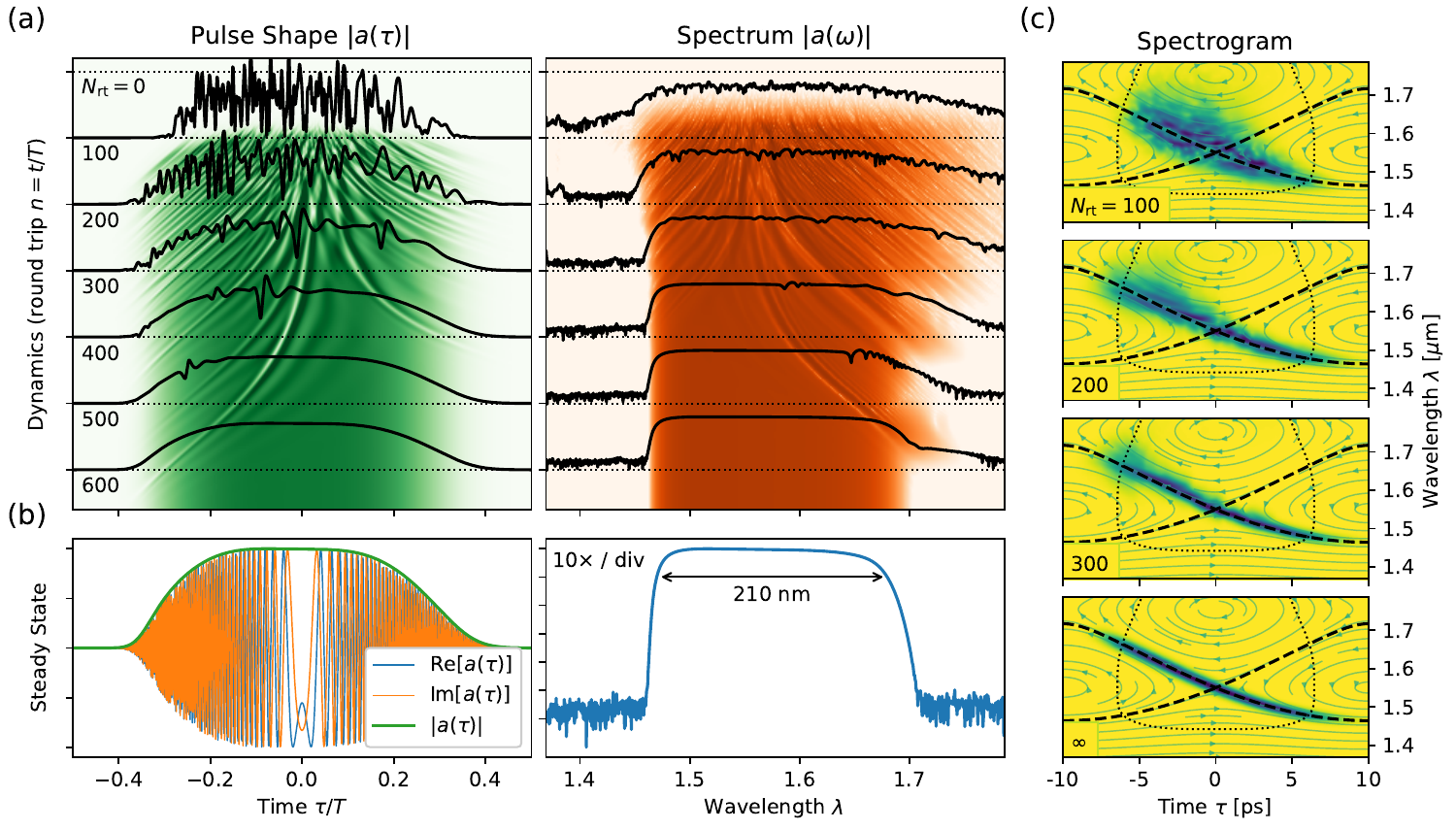}
\caption{Simulation of comb formation. (a) Transient and (b) steady-state pulse shape $a(t)$ and spectrum $a(\omega)$. (c) Evolution of spectrogram of $a(\tau)$, showing stabilization of FMCW waveform.  Settings: $f_{\rm rep} = 50$~GHz, $\phi_a = 0.4$, $\phi_p = 1.6$, DCF = 2.}
\label{fig:f4}
\end{center}
\end{figure*}

To study the QAM-OPO in more detail, consider the specific benchmark configuration of Table~\ref{tab:t1}, designed to have realistic experimental parameters, and which we will revisit over the course of this paper.  The parameters considered here correspond to a cavity containing both guided-wave and free-space sections. The discrete OPA and EOM segments (1-cm each) are assumed to be either bulk of weakly-guiding LiNbO$_3$, where the material dispersion dominates over geometric (waveguide) dispersion. The frameworks and dynamical regimes studied here are readily extended to integrated platforms, such as thin-film lithium niobate \cite{zhu2021integrated}, where the dispersion relations of each section may be drastically altered by the waveguide geometry \cite{jankowski2022quasi}.  
The out-coupling loss is 20\%, corresponding to a round-trip loss of 1~dB.  We also include the option of dispersion compensation factor (DCF, default unity), modeled as an auxiliary dispersion element that reduces the GDD $\delta_2 \equiv \delta''(\omega_a)$ by the factor $\delta_2 \rightarrow \delta_2/\text{DCF}$.  The default dispersion compensation material is SMF28 fiber unless otherwise specified.  The EOM frequency $f_{\rm rep}$ and driving phases $\phi_a, \phi_p$ have no default values.  The OPA bandwidth also depends on the pumping condition; for the broadest combs (see Sec.~\ref{sec:bw}), we primarily consider the case of signal-idler group-velocity (GV)-matching, Table~\ref{tab:t2}.  These defaults in mind, Fig.~\ref{fig:f3} provides an illustration of the phase-space model Eq.~(\ref{eq:2-char}), with the dynamics of $\tau$ and $\omega$ (plotted as $\lambda = 2\pi c/\omega$) plotted as streamlines, and the dynamics of $A$ plotted as a contour map.  

Fig.~\ref{fig:f4} illustrates the dynamics of pulse formation in this QAM-OPO.  Here we assume instantaneous pump turn-on at round trip $n = t/T = 0$, where the field initially resembles parametrically amplified vacuum noise.  This leads to an irregular pulse intensity $P(\tau) = |a(\tau)|^2$ and power spectrum $P(\omega) = |a(\omega)|^2$.  However, over the course of the several hundred round-trips in Fig.~\ref{fig:f4}(a), these ``glitches'' are pushed away from the center of the pulse, where they are eventually suppressed by the intracavity IM, leaving behind a smooth waveform and spectrum.  The steady-state, plotted in Fig.~\ref{fig:f4}(b), is a rapidly oscillatory waveform with a slowly-varying amplitude, consistent with the FMCW ansatz of Eq.~(\ref{eq:2-fmcw}).  In the frequency domain, this corresponds to a flat spectrum that covers (in this case) the entire gain bandwidth of the OPA crystal, which for C-Band GV matching is around 200~nm (approximately 500 comb lines at $f_{\rm rep} = 50$~GHz).  It is also instructive to look at the spectrogram of the QAM-OPO signal, plotted in Fig.~\ref{fig:f4}(c), which highlights the utility of the phase-space model of Eqs.~(\ref{eq:2-char}) in understanding the pulse dynamics, namely, that an initial noisy distribution of power is amplified and then ``sheared'' by the EOM chirp and GDD, following the phase portrait of Fig.~\ref{fig:f3}, leaving a steady-state pulse centered at the saddle point and emanating out to the edges of the gain window.  This evolution is deterministic and ``turn-key'', so the same comb would be formed with a gradual pump turn-on, in contrast to bright-soliton Kerr combs \cite{herr2014temporal} or doubly-resonant / pump-resonant OPOs \cite{lu2021ultralow, chen2021efficient}, where dynamical effects (and often thermal locking \cite{gray2020thermo}) are key to achieving the desired end state.

\subsection{Steady-State Solution}
\label{sec:2-ss}

Integrating Eqs.~(\ref{eq:2-char}) for $(\omega, \tau)$, we observe that the characteristic curves follow a {\it conservative} dynamical flow, with the round-trip phase being a constant:
\beq
	\delta(\omega) + \phi(\tau) = C \label{eq:2-constphase}
\eeq
Intuitively, for a signal to be stable from one round trip to the next, the total phase accumulated (GDD phase plus EOM phase) must the same everywhere in the pulse.  This ``constant-phase'' model allows us to quickly find good analytic approximations for the comb bandwidth and spectrum, without solving any dynamical models.  

Stable operation of the QAM-OPO is obtained when the gain maximum overlaps with a saddle point on the phase portrait as in Fig.~\ref{fig:f3}, and the steady-state solution branches out from that point along the separatrix, the constant-phase curve that crosses the saddle point (dashed lines in Fig.~\ref{fig:f4}(c)).  Like any fixed point, this saddle point is located at:
\beq
	\phi'(\tau) = 0,\ \ \ 
	\delta'(\omega) = 0
\eeq
The first condition places the saddle at the maximum (or minimum) of the EOM phase, while the second condition physically corresponds to matching the cavity FSR to the EOM frequency, $\text{FSR} = f_{\rm rep}$.  Since the FSR is a function of $\omega$ through cavity dispersion, this sets the value of $\omega$ at the saddle point.

Since these dynamics are not affected by a global phase shift, we can shift $\phi(\tau)$ and $\delta(\omega)$ so that $\phi = \delta = 0$ at the saddle point, in which case Eq.~(\ref{eq:2-constphase}) gives $C = 0$ for the separatrix.  Using Eq.~(\ref{eq:2-constphase}), we can solve for the pulse chirp $\omega(\tau)$ and amplitude $A(\tau)$:
\begin{align}
	\omega(\tau) & = \delta^{-1}(-\phi(\tau)) \label{eq:2-dw} \\
	\delta'(\omega(\tau))\frac{\d A}{\d \tau} & = \frac{1}{2}\bigl[g(A, \omega(\tau)) - \alpha(\tau)\bigr] A \label{eq:2-da2}
\end{align}
Here, it is instructive to consider the quasi-static limit, where the left-hand side can be ignored and the quasi-static field, denoted $A_0$ is obtained by solving $g(A_0, \omega(\tau)) = \alpha(\tau)$.  This is solved using Eq.~(\ref{eq:2-gnorm}):
\beq
	A_0 = \frac{1}{\kappa L} \sqrt{\sinc^{-1}\Bigl(\frac{\sqrt{\alpha(t)/\alpha_0}}{p}\Bigr)^2 - \Bigl(\frac{\Delta\beta[\omega(t)] L}{2}\Bigr)^2}
\eeq
This has a sharp cutoff when $\sinc^2\bigl(\tfrac12\Delta\beta[\omega(t)]L\bigr) = \alpha(t)/(\alpha_0p^2)$; beyond this point, net gain is not possible as the OPO is locally below threshold, and $A_0 = 0$.

\subsection{Inverted Pendulum Model}
\label{sec:2-pend}

\begin{figure}[tbp]
\begin{center}
\includegraphics[width=1.00\columnwidth]{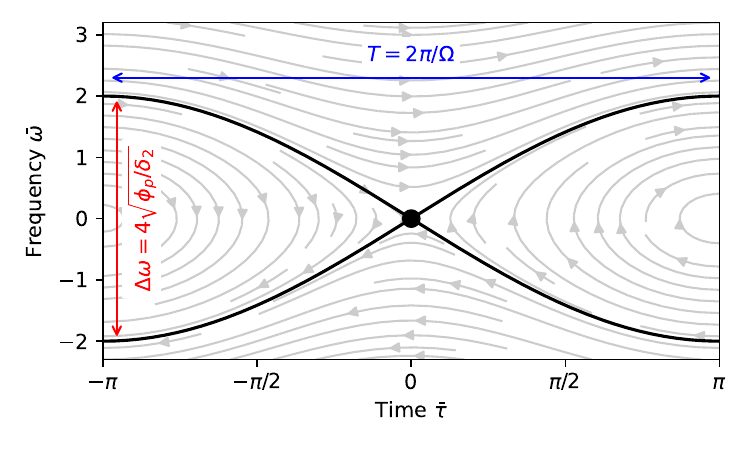}
\caption{Phase portrait of the inverted pendulum in normalized coordinates.}
\label{fig:f5}
\end{center}
\end{figure}

A useful simplification is to (i) truncate dispersion at second order, $\delta(\omega) = \tfrac12 \delta_2 \omega^2$, and (ii) assume that the EOM is driven by a single frequency, i.e.\ that Eq.~(\ref{eq:2-rp}) holds, equivalent to $\phi(\tau) = -2\phi_p \sin^2(\Omega\tau/2)$.  In this case, Eqs.~(\ref{eq:2-char}) takes the form:
\beq
	\frac{\d\tau}{\d n}   = \delta_2\omega,\ \ \ 
	\frac{\d\omega}{\d n} =  \phi_p \Omega \sin(\Omega\tau) \label{eq:2-pend}
\eeq
Note the resemblance of the $(\tau, \omega)$ equations to the inverted pendulum $\dot{x} = p/m, \dot{p} = m g \sin(x/\ell)$ (for a pendulum of mass $m$, length $\ell$, in gravitational field $g$).  The phase portrait of the pendulum is sketched in Fig.~\ref{fig:f5}.  The time-frequency chirp, computed with Eq.~(\ref{eq:2-dw}), is:
\beq
	\omega(t) = 2\sqrt{\phi_p/\delta_2} \sin(\Omega t/2) \label{eq:2-tfc}
\eeq
which gives an upper bound on the comb bandwidth
\beq
	\Delta\omega_{\rm BW} = 4\sqrt{\phi_p/\delta_2},\ \ \ 
	\Delta\lambda_{\rm BW} = \frac{2\lambda^2 \sqrt{\phi_p/\delta_2}}{\pi c} \label{eq:2-dwmax}
\eeq
which is twice the maximum excursion of $\omega(t)$ from the separatrix.  For the example in Figs.~\ref{fig:f3}-\ref{fig:f4}, $\delta_2 \approx 1000~\text{fs}^2$ (2-cm LiNbO$_3$ with $2\times$ DCF) and $\phi_p = 1.6$, so Eq.~(\ref{eq:2-dwmax}) predicts a 200-nm comb.  The actual bandwidth is slightly larger due to higher-order dispersion, which deviates from the pendulum model.

Finally, we can normalize the pendulum model by converting to dimensionless coordinates:
\beq
	\tau = \Omega^{-1}\bar{\tau},\ \ \ 
	\omega = \underbrace{\sqrt{\phi_p/\delta_2}}_{\xi_\omega}\,\bar{\omega},\ \ \ 
	n = \underbrace{\vphantom{\sqrt{\phi_p/\delta_2}}(\Omega\sqrt{\delta_2\phi_p})^{-1}}_{\xi_n} \bar{n} \label{eq:2-norm}
\eeq
This normalization indicates the relevant frequency- and time-scale of the OPO: to order of magnitude, $1/\Omega$ governs the pulse size, $\xi_\omega$ the comb bandwidth, and $\xi_n$ the number of round trips required to build up a stable comb.  For the parameters used above, $\xi_\omega/2\pi = 6.4$~THz (50~nm) and $\xi_n = 80$.  In normalized variables, Eqs.~(\ref{eq:2-pend}) become:
\beq
	\frac{\d\bar{\tau}}{\d\bar{n}}     =  \bar{\omega},\ \ \ 
	\frac{\d\bar{\omega}}{\d\bar{n}}   =  \sin(\bar{\tau}),\ \ \ 
	\frac{\d A}{\d\bar{n}}             =  \frac{\xi_n}{2} \bigl[g(A,\omega)-\alpha(\bar{\tau})\bigr] A \label{eq:2-pendr}
\eeq
which is the inverted pendulum model.  The general solutions are elliptic functions, which trace out constant-phase curves $\bar{\omega}^2 - (2\sin(\bar{\tau}/2))^2 = C$, with the separatrix ($C = 0$) having the simple expression $\bar{\omega} = \pm 2\sin(\bar\tau/2)$ (compare Eq.~(\ref{eq:2-tfc})), giving a bandwidth bound of $\Delta\bar{\omega} = 4$ in dimensionless units (compare Eq.~(\ref{eq:2-dwmax})).

\section{Stability Analysis and Limitations} \label{sec:stability}

\begin{figure*}[p]
\begin{center}
\includegraphics[width=1.00\textwidth]{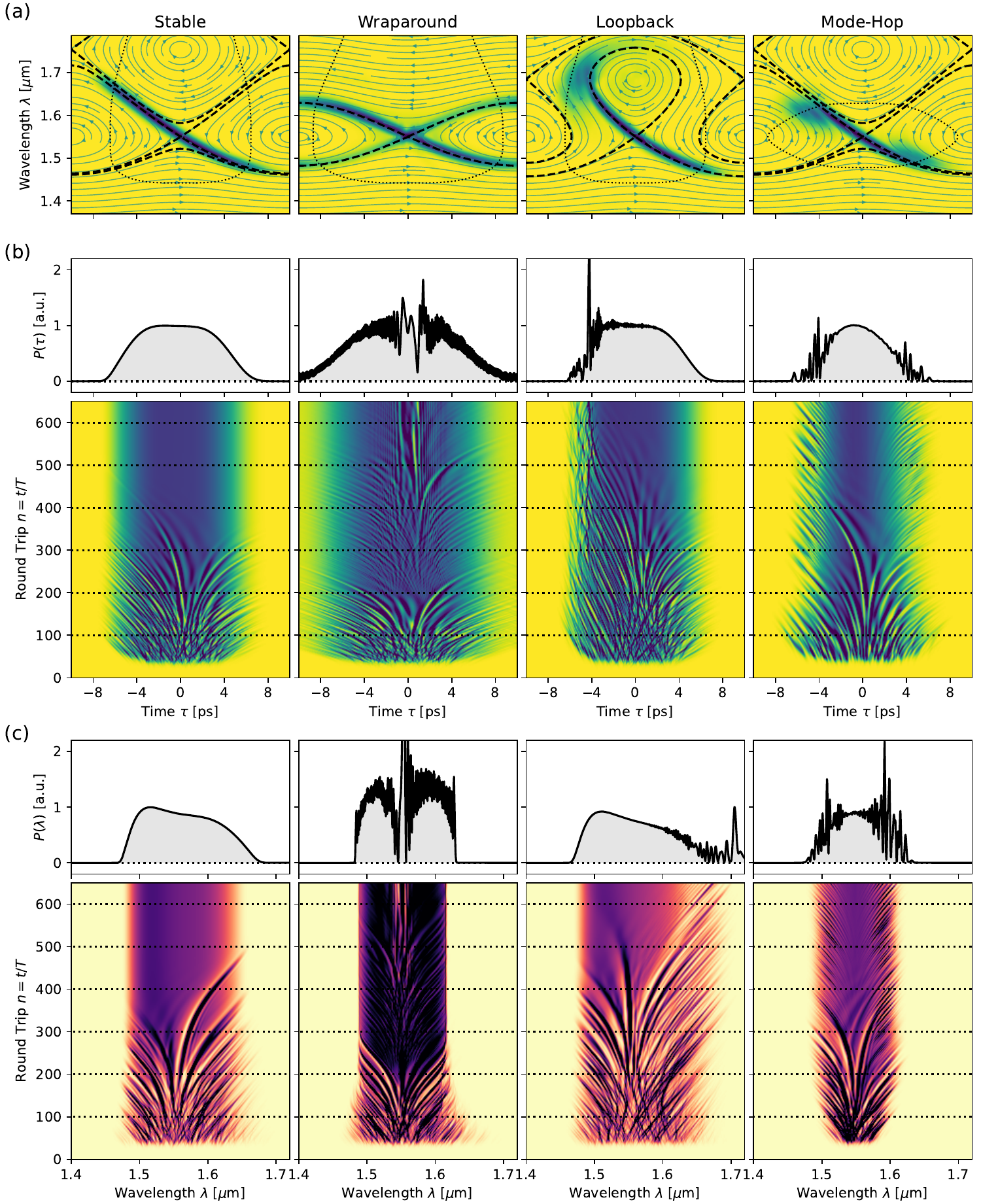}
\caption{QAM-OPO instabilities.  Comparison of stable dynamics, wraparound instability, loopback instability, and mode-hopping instability.  (a) Pulse spectrogram. (b) Power $P(\tau) = |a(\tau)|^2$.  (c) Power spectrum $P(\omega) = |\tilde{a}(\omega)|^2$.  Conditions for the simulations are as follows: (Stable) $f_{\rm rep} = 50$~GHz, DCF = 2, $\phi_p = 1.6$~rad, $\phi_a = 0.4$~rad; (Wraparound) DCF = 1, $\phi_a = 0.3$~rad; (Loopback) DCF = 2.7; (Mode-hop) $\lambda_c = 0.912~\mu$m, $\Lambda = 26.66~\mu$m, compare Table~\ref{tab:t2}.}
\label{fig:3-f5}
\end{center}
\end{figure*}

\begin{figure*}[t!]
\begin{center}
\includegraphics[width=1.00\textwidth]{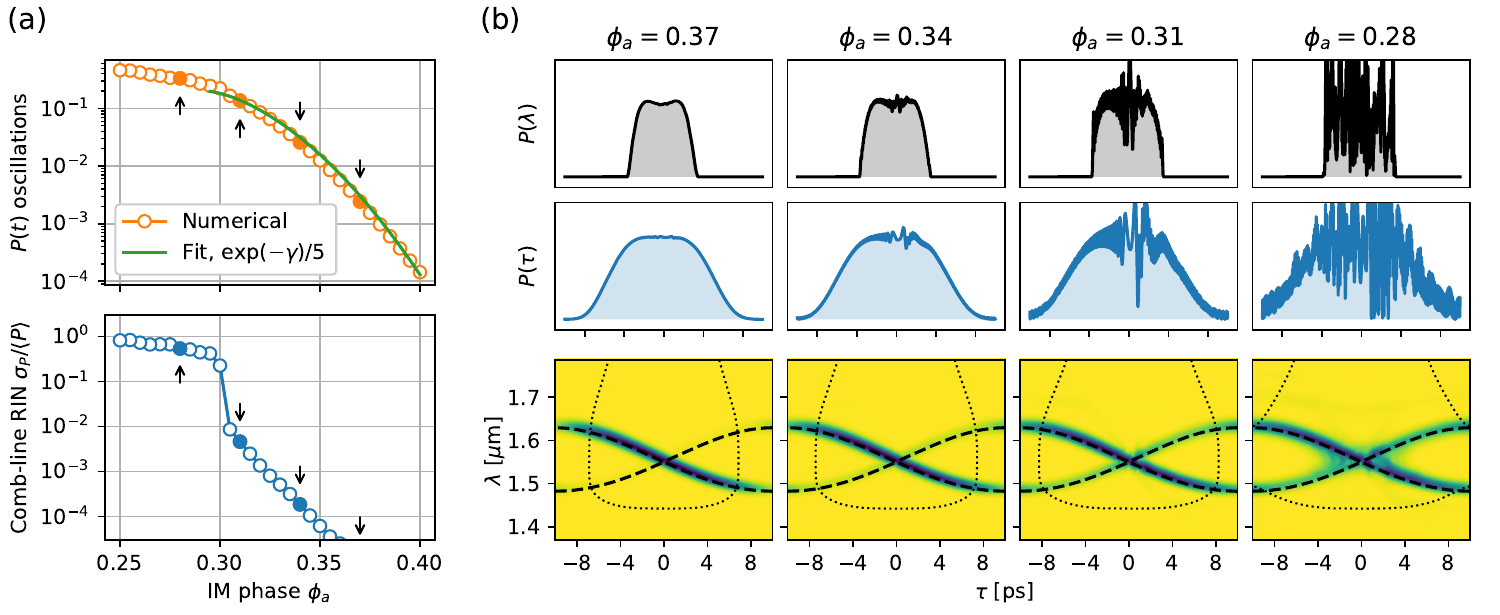}
\caption{Wraparound Instability.  (a) Oscillations in power $P(t)$ and comb-line RIN noise $\sigma_P / \langle P \rangle$ as a function of IM phase $\phi_a$.  (b) Pulse waveform, spectrum, and spectrogram for a range of $\phi_a$.  Parameters: $f_0 = 50~$GHz, $\phi_p = 1.6$~rad.}
\label{fig:3-f6}
\end{center}
\end{figure*}

Operating the QAM-OPO involves the careful art of navigating a landscape of potential instabilities.  As shown in the previous section, stable comb formation is robust and deterministic under the right conditions; however, in certain regimes, instabilities can disrupt the FMCW solution and lead to unstable, noisy combs that are not useful in most applications.  The three most important instabilities, depicted in Fig.~\ref{fig:3-f5}, are (1) the wraparound instability, (2) the loopback instability, and (3) the mode-hop instability.  This section studies these instabilities and the conditions that give rise to them.

\subsection{Wraparound Instability}
\label{sec:3-wrap}

Dividing the $A$ and $\bar{\tau}$ equations in Eq.~(\ref{eq:2-pendr}), we obtain an equation for $A(\tau)$, which we wish to evaluate at the separatrix:
\beq
	\frac{\d A}{\d\bar{\tau}} = \frac{\xi_n}{2\bar{\omega}} (g-\alpha) A
	\stackrel{\rm sep}{\longrightarrow} \frac{\xi_n}{4\sin(\bar{\tau}/2)} (g-\alpha) A \label{eq:3-dap}
\eeq
A first-approximation to Eq.~(\ref{eq:3-dap}) is to set the right-side term $(g-\alpha)A$ to zero, since $\xi_n \gg 1$ and $\d A/\d\bar{\tau} \sim O(1)$ in realistic regimes.  This leads to the solution:
\beq
	\begin{dcases}
		g(A, \omega) = \alpha(\tau) & \text{(above threshold)} \\
		A = 0 & \text{(below threshold)}
	\end{dcases} \label{eq:3-qs}
\eeq
The intracavity IM divides phase space into a ``gain'' region centered around $\bar{\tau} = 0$ and a ``loss'' region centered around $\bar{\tau} = \pi$ (Fig.~\ref{fig:f3}).  This is designed to clip the tails of the pulse and prevent wraparound instability.  Wraparound is suppressed completely in the limit $\xi_n \rightarrow \infty$, where Eq.~(\ref{eq:3-qs}) is exact.  However, in realistic situations, the GVD term cannot be completely ignored; and the tails of the pulse will still leak into the below-threshold region with an amplitude that scale as
\beq
	A(\bar{\tau}) \propto \exp\Bigl(\xi_t \int_{\bar{\tau}_0}^{\bar{\tau}} \frac{1}{4\sin(\bar{\tau}/2)} \bigl[g - \alpha(\bar{\tau})\bigr]\d\bar{\tau}\Bigr) \label{eq:3-aexp}
\eeq
where $\bar{\tau}_\pm$ are the boundaries of the loss region ($g - \alpha \leq 0$ for $\bar{\tau} \in [\bar{\tau}_-, \bar{\tau}_+]~(\text{mod}\,2\pi)$).  Since $g < \alpha$ in the loss region, Eq.~(\ref{eq:3-aexp}) predicts $A(\bar{\tau})$ will be exponentially attenuated.  Wraparound still occurs, but only at a controllably small amplitude.

To estimate analytically the magnitude of any residual wraparound, we make two approximations: (1) assume the comb is well within the OPA's phase-matching bandwidth, so that the gain given by Eq.~(\ref{eq:2-gnorm}) reduces to $g \rightarrow p^2\alpha_0$, which independent of $\omega$, and (2) assume the amplitude modulation is sufficiently weak that the expression $\alpha(\tau) = \alpha_0 + 2 \log[\sec(\rho(\tau))]$ can be expanded to leading order in $\phi_a$.  Thus we have
\begin{align}
	& (g - \alpha) \rightarrow (p^2 - 1)\alpha_0 - 4\phi_a^2 \sin^4(\bar{\tau}/2) \\
	& \tau_- = 2\sin^{-1}\Bigl[\Bigl(\frac{(p^2-1)\alpha_0}{4\phi_a^2}\Bigr)^{1/4}\Bigr],\ \ \ 
	\tau_+ = 2\pi - \tau_-
\end{align}
The wrapped-around field has the amplitude:
\begin{align}
	& A_{\rm wrap} \propto \exp\Bigl(\xi_t \int_{\bar{\tau}_-}^{\bar{\tau}_+} \frac{(p^2 - 1)\alpha_0 - \phi_a^2 \sin^4(\bar{\tau}/2)}{4\sin(\bar{\tau}/2)} \d\bar{\tau}\Bigr) \nonumber \\
	& \approx \exp\Bigl(-\underbrace{\frac{\sqrt{2}}{3} \xi_t (p^2-1)\alpha_0 \Bigl[\frac{4\phi_a^2}{(p^2-1)\alpha_0} - 1\Bigr]^{3/2}}_{\gamma}\Bigr) \label{eq:3-egam}
\end{align}
where the second line of Eq.~(\ref{eq:3-egam}) is taken by expanding the integral to leading order in $\phi_a^2$ around $\phi_a^2 = (p^2 - 1)\alpha_0$.

\begin{figure*}[t!]
\begin{center}
\includegraphics[width=1.00\textwidth]{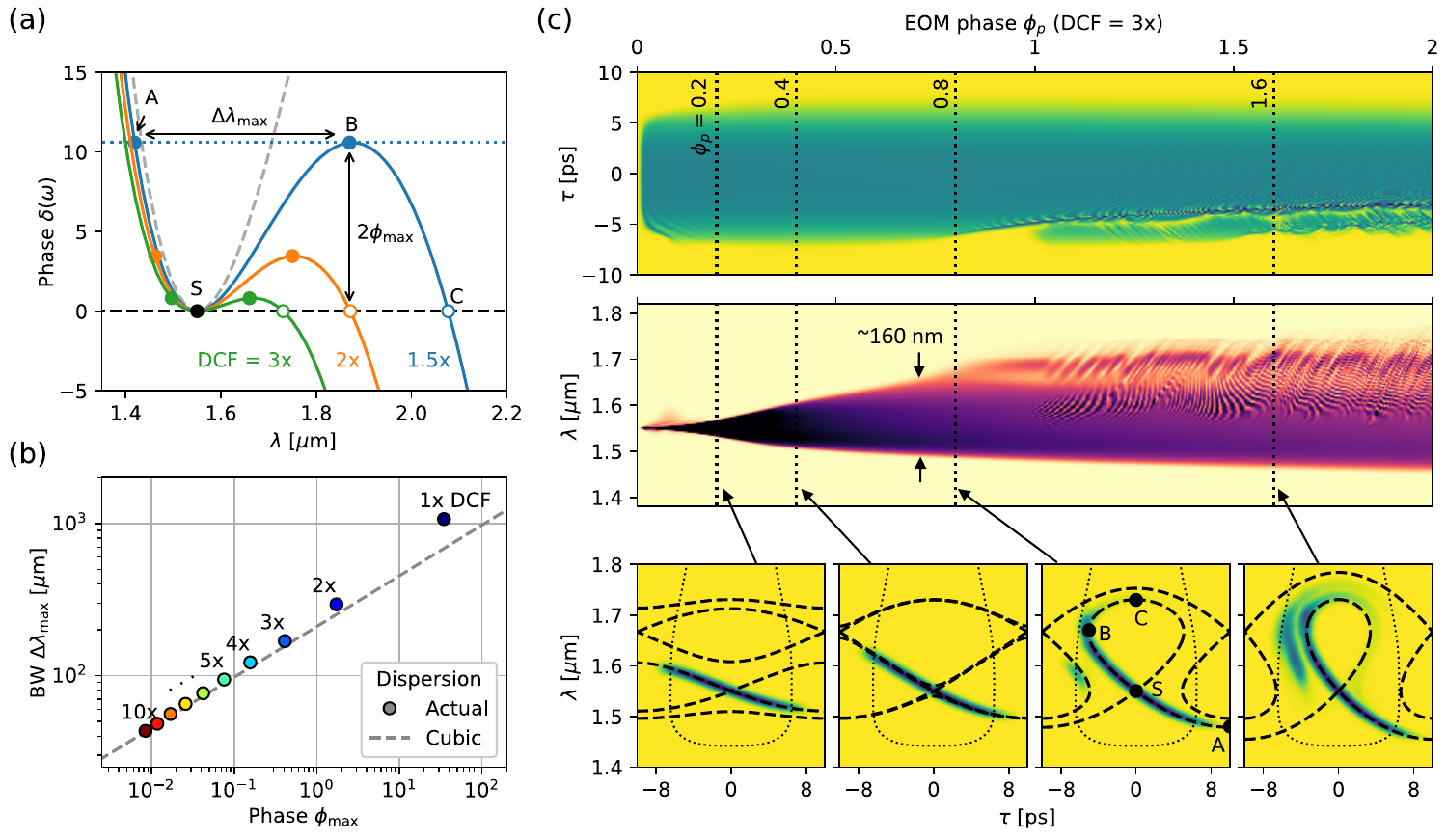} %
\caption{Loopback instability.  (a) TOD causes the dispersive phase curve $\delta(\omega)$ to bend over, so the PM drive phase and comb bandwidth are limited.  (b) Calculated $\phi_{\rm max}$ and $\Delta\lambda_{\rm max}$ as a function of dispersion compensation, as compared to the analytic result Rq.~(\ref{eq:3-lpmax}).  (c) Simulation of onset of instability.  Phase drive is ramped from $\phi_p = 0$ to $2$ over 2000 round trips.  Parameters: $f_0 = 50$~GHz, $\phi_a = 0.4$~rad, DCF = $3\times$.}
\label{fig:3-f7}
\end{center}
\end{figure*}

Fig.~\ref{fig:3-f6} illustrates the onset of wraparound stability and the importance of IM in maintaining a stable comb.  Here, we slowly reduce the IM phase in an QAM-OPO that initially shows stable comb formation (Fig.~\ref{fig:3-f6}(b), left column).  As $\phi_a$ is reduced, the loss region narrows and the field begins to wrap around, interfering with itself and producing the fringes in $P(t)$ as observed in the middle columns of Fig.~\ref{fig:3-f6}(b).  The amplitude of these oscillations can be quantified by high-pass filtering $P(t) \rightarrow F \otimes P$ and comparing the norm to that of the unfiltered function; the resulting ratio $\int{|F \otimes P|^2\d t} / \int{|P|^2\d t}$ is plotted in Fig.~\ref{fig:3-f6}(a).  We observe a steady increase in the oscillations, which correlate closely with the tunneling $e^{-\gamma}$, where $\gamma$ is the tunneling amplitude obtained in Eq.~(\ref{eq:3-egam}).

Small amounts of wraparound lead to oscillations, but the waveform is still stable.  However, at around $\phi_a = 0.3$, an abrupt transition occurs and the comb becomes chaotic.  This happens because the wraparound amplitude is so large that nonlinear effects destabilize the comb at its saddle point.  This transition is evident in the relative intensity noise (RIN) of individual comb lines, plotted in Fig.~\ref{fig:3-f6}(a), which shows a sharp jump at $\phi_a = 0.3$ and plateaus near $O(1)$.

\subsection{Loopback Instability}
\label{sec:3-loop}

In any model with dispersion truncated to second order, one can always increase the comb bandwidth by scaling the relevant parameters -- in this case phase $\phi_p$ and GDD $\delta_2$.  In practice, just like in Kerr combs, third-order dispersion (TOD) limits the bandwidth that can be practically achieved.  In the presence of TOD, the bandwidth of a comb is limited by a change in the sign of $\beta_2$ at the zero-dispersion wavelength $\lambda_{\rm zdw}$.  This leads to dispersive-wave emission (also referred to as Cherenkov rediation) in both solitons \cite{wai1990radiations, akhmediev1995cherenkov} and Kerr combs \cite{jang2014observation, brasch2016photonic}, and for the QAM-OPO, gives rise to {\it loopback instability}, where the separatrix loops back on itself, so that $\omega(\tau)$ loses its monotonicity.  In this case, the FMCW assumption no longer holds, since in the region of the loop the field is the sum of two frequencies, which beat against each other.  Loopback instability is problematic for two reasons: (1) thanks to the loop, a fraction of the gain medium goes undepleted, which leads to uncontrolled amplification of vacuum fluctuations, and (2) if a significant fraction of the field traverses the loop, it can re-emerge at the saddle point and destabilize the comb.

Recall that the separatrices are curves of constant phase: $\delta(\omega) + \phi(\tau) = 0$.  If our EOM phase spans the range $[0, -2\phi_p]$, then to maintain the constant phase, the separatrices compensatingly span the range $\delta \in [0, +2\phi_p]$.  This can be visualized by plotting the $\delta(\omega)$ as a potential well (Fig.~\ref{fig:3-f7}(a)): the comb forms in the region $\delta(\omega) < 2\phi_p$.  In the quadratic case (Sec.~\ref{sec:2-pend}), this yields $|\omega| < 2\sqrt{\phi_p/\delta_2}$ for a total bandwidth of $\Delta\omega_{\rm BW} = 4\sqrt{\phi_p/\delta_2}$.

However, in the presence of higher-order dispersion, the potential well in Fig.~\ref{fig:3-f7}(a) bends over.  As a result, there is a maximum drive phase $\phi_{\rm max}$, which supports a comb of maximum bandwidth $\Delta\omega_{\rm max} = |\omega_B - \omega_A|$ between point $B$ (the local maximum of $\delta(\omega)$) and its equipotential $A$.  Since $B$ is a maximum of $\delta(\omega)$, $\dot{\tau} \propto \delta'(\omega_B) = 0$, meaning that $\dot{\tau}$ changes sign and the characteristic ``loops back'' on itself.  This is the origin of the loopback instability.

To good approximation, we can truncate dispersion at third order, so that the dispersive phase takes the form $\delta(\omega) = \tfrac12 \delta_2\omega^2 + \tfrac16 \omega^3$.  We find:
\beq
	\omega_A = \delta_2/\delta_3,\ \ \ 
	\omega_B = -2\delta_2/\delta_3,\ \ \ 
	\omega_C = -3\delta_2/\delta_3
\eeq
The resulting maximum bandwidth and EOM phase are:
\beq
	\phi_{\rm max} = \frac{\delta_2^3}{3\delta_3^2},\ \ \ 
	\Delta\omega_{\rm max} = \frac{3\delta_2}{\delta_3} = 3 \Bigl(\frac{3\phi_{\rm max}}{\delta_3}\Bigr)^{1/3} \label{eq:3-lpmax}
\eeq
Loopback instability will become problematic if significant amounts of dispersion compensation are used, since usually dispersion is only compensated to leading order.  In the quadratic model, dispersion compensation should improve the bandwidth of the comb due to the scaling $\Delta\omega \propto 1/\sqrt{\delta_2}$.  However, since $\phi_{\rm max} \propto \delta_2^3$ per Eq.~(\ref{eq:3-lpmax}), the overall effect will be detrimental unless $\phi_{\rm max}$ is large compared to the modulator drive voltage.

To give a sense of perspective, Fig.~\ref{fig:3-f7}(b) plots $\phi_{\rm max}$ and $\Delta\lambda_{\rm max} = (\lambda^2 /2\pi c)\Delta\omega_{\rm max}$ for EO Comb OPOs with various degrees of dispersion compensation (parameters from Table~\ref{tab:t1} used, uncompensated $\delta_2 = 2000~\text{fs}^2$, $\delta_3 = 6500~\text{fs}^3$).  Dispersion compensation is performed using standard SMF28 fiber ($\beta_2 = -20~\text{fs}^2/\text{mm}$, $\beta_3 = 122~\text{fs}^3/\text{mm}$) to reduce the GVD by the desired factor.  We see that without dispersion compensation, $\phi_{\rm max} \sim 50$, which is much larger than the drive phase of realistic modulators; however, even a modest $2$--$3\times$ GVD reduction reduces $\phi_{\rm max}$ to around unity, so loopback instability cannot be ignored.  The small drive phases and correspondingly narrow bandwidths suggest that large amounts of compensation $\text{DCF} > 10\times$ will be unhelpful in this case.

Fig.~\ref{fig:3-f7}(c) numerically investigates the onset of instability by adiabatically sweeping the EOM phase from $\phi_p = 0$ to $4$.  Here, we have used DCF = 3, so that $\delta_2 = 650~\text{fs}^2$ and $\delta_3 = 14000~\text{fs}^3$; from Eq.~(\ref{eq:3-lpmax}), the corresponding phase and bandwidth limits are $\Delta\phi_{\rm max} = 0.9$, $\Delta\lambda_{\rm max} = 170$~nm.  This is roughly in line with the numerical results, where we observe stable comb formation up until about $\phi = 1.5$, after which the spectrograms show significant loopback and spontaneous amplification of vacuum noise.

\subsection{Mode-Hop Instability}
\label{sec:3-modehop}

Finally, the pulse can be disrupted by mode-hopping phenomena.  To understand this instability, recall that the QAM-OPO outputs FMCW pulses, and the dynamics are governed by quasi-CW physics.  As a result, CW OPO phenomena shed important insight into the stability of the QAM-OPO pulse.  In particular, CW OPO physics is a classic case of multimode gain with a common pump, where a large number of (longitudinal) modes compete for limited pump gain, and the one with the largest gain (i.e.\ phase-matched) wins out and oscillates \cite{siegman1986lasers}.  Therefore, if the OPO oscillates at a sub-optimal (phase-mismatched) wavelength, this state is at best metastable, and the system may ``mode hop'' to a higher-gain signal mode.

Mode hopping is a problem for the QAM-OPO because, for finite phase-matching bandwidth, it is not possible for all comb lines to experience the same gain.  The tails of the FMCW pulse are necessarily detuned from the center wavelength, so if the center of the comb is phase-matched, the tails will be phase-mismatched and experience less gain, and are thus subject to mode-hopping.  If the quasi-CW picture were exact, this would stabilize any comb.  Fortunately, the phase-space dynamics of Sec.~\ref{sec:2-fmcw} only allow this instability to act for a limited time before being pushed away from the gain region and damped, and this saves the QAM-OPO.  But we still need to study the instability and understand when it occurs.

\begin{figure*}[t!]
\begin{center}
\includegraphics[width=1.00\textwidth]{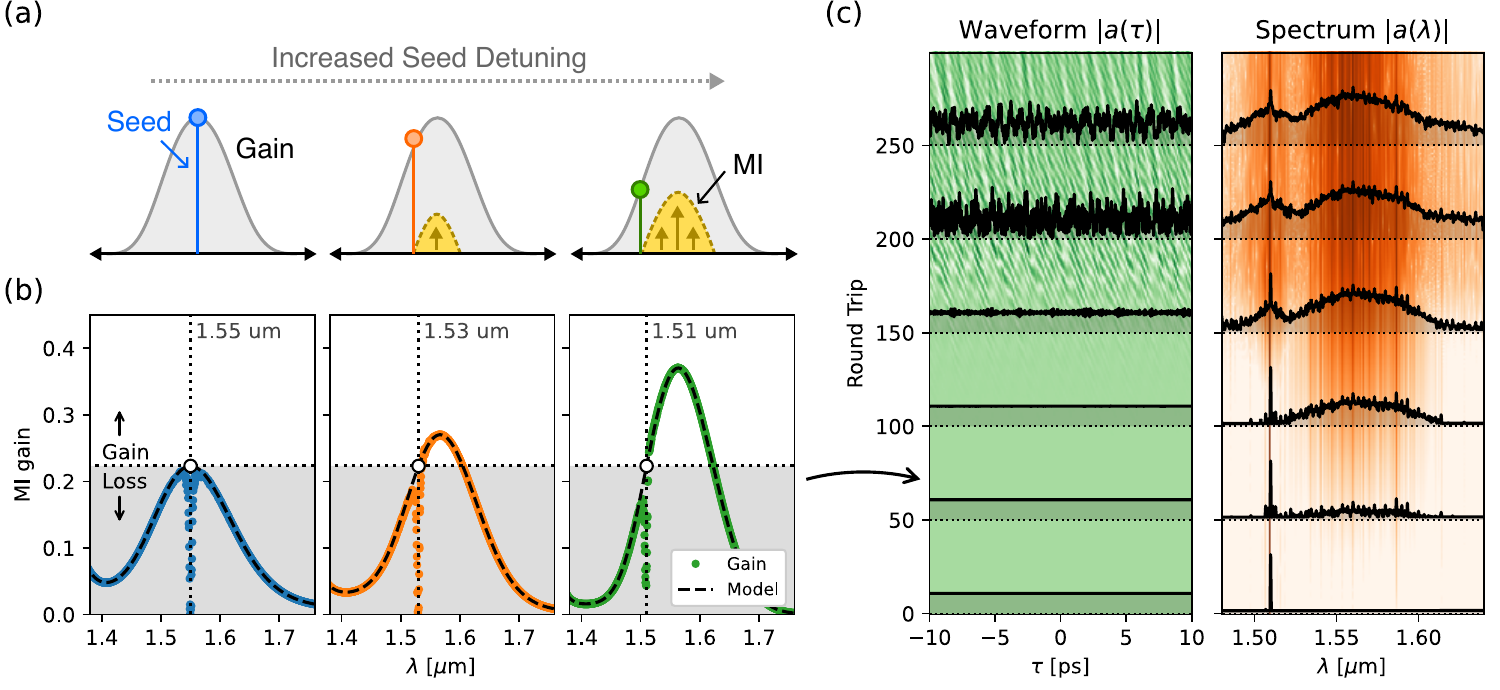}
\caption{Mode hopping of a CW field.  (a) Illustration of modulation instability when an OPO is oscillating at a non-optimal signal wavelength.  (b) MI gain $g_{\rm MI}$ for an OPO phase-matched to 1.55~$\mu$m and seeded at different signal wavelengths.  Colored dots are full numerical simulations; dashed line is the reduced model Eq.~(\ref{eq:3-dp3}).  (c) Numerical evolution of a CW OPO initialized to a 1.51~$\mu$m signal, showing the growth of MI modes that disrupt the original CW field.  Simulation: Table~\ref{tab:t1} parameters with $f_{\rm rep} = 50$~GHz and no modulation, $\phi_a = \phi_c = 0$.}
\label{fig:3-f8}
\end{center}
\end{figure*}

To begin, we analyze mode-hopping in a CW OPO by perturbing the field equations Eqs.~(\ref{eq:2-d}) (written in frequency basis, see Eqs.~(\ref{eq:a-df})) about a steady-state solution:
\begin{subequations}\begin{align}
	\delta\dot{a}_m & = i \beta_{a,m} \delta a_m + i\kappa \sum_n (\delta c_{m+n} b_n^* + c_{m+n}\delta b_n^*) \label{eq:3-da-p} \\
	\delta\dot{b}_m & = i \beta_{b,m} \delta b_m + i\kappa \sum_n (\delta c_{m+n} a_n^* + c_{m+n}\delta a_n^*) \label{eq:3-db-p} \\
	\delta\dot{c}_m & = i \beta_{c,m} \delta c_m + i\kappa \sum_n (\delta a_n b_{m-n} + a_n\delta b_{m-n}) \label{eq:3-dc-p}
\end{align}\label{eq:3-d-p}\end{subequations}
Mode-hopping begins as a modulation instability on top of a CW background (seed) field (Fig.~\ref{fig:3-f8}(a)).  This background is obtained by solving for the steady-state equations (numerically or analytically) in Appendix~\ref{sec:a-opo}.  Without loss of generality, we center the frequency window so that the CW background lies in the $m = 0$ mode $(a_0, b_0, c_0)$.  Eqs.~(\ref{eq:3-d-p}) then reduce to:
\begin{subequations}\begin{align}
	\delta\dot{a}_m & = i \beta_{a,m} \delta a_m + i\kappa (b_0^* \delta c_m + c_0 \delta b_{-m}^*) \label{eq:3-dap2} \\
	\delta\dot{b}_m & = i \beta_{b,m} \delta b_m + i\kappa (a_0^* \delta c_m + c_0 \delta a_{-m}^*) \label{eq:3-dbp2} \\
	\delta\dot{c}_m & = i \beta_{c,m} \delta c_m + i\kappa (b_0^* \delta a_m + a_0 \delta b_m^*) \label{eq:3-dcp2}
\end{align}\label{eq:3-dp2}\end{subequations}
For each $m \neq 0$, which corresponds to a modulation frequency $m\Omega$, Eqs.~(\ref{eq:3-dp2}) are a linear system of six ODEs for the variables $(\delta a_{\pm m}, \delta b_{\pm m}, \delta c_{\pm m})$.  By solving these ODEs numerically over a single OPO round-trip, one can calculate the transition matrix and determine whether the perturbations will grow.  The dynamics simplify when the pump and signal have significant group-velocity mismatch (which is usually the case, since we would rather dispersion-engineer the OPO to match signal and idler group velocities, and matching three group velocities is very difficult).  In this case, the pump-mode perturbation $\delta c_m$ is highly phase-mismatched due to a large $\beta_{c,m}$, and therefore it can be removed from the dynamics, yielding a $2\times2$ system for the variables $(\delta a_m, \delta b_m)$:
\beq
	\frac{\d}{\d z} \begin{bmatrix} \delta a_m \\ \delta b_{-m}^* \end{bmatrix}
	= \begin{bmatrix} i \beta_{a,m} & \kappa c_0(z) \\ \kappa c_0(z)^* & -i\beta_{b,-m} \end{bmatrix}
	\begin{bmatrix} \delta a_m \\ \delta b_{-m}^* \end{bmatrix} \label{eq:3-dp3}
\eeq
Solving Eq.~(\ref{eq:3-dp3}) gives the modulation instability (MI) gain $g_{\rm MI}(m\Omega) = 2 \log |a_{m}(L)/a_{m,\rm}(0)|$.  In Fig.~\ref{fig:3-f8}(b), we calculate the MI gain for seed fields of different detunings from the optimal phase-matching signal (1.55~$\mu$m), using both Eq.~(\ref{eq:3-dp3}) and full numerical simulations.  We see that Eq.~(\ref{eq:3-dp3}) is accurate for all frequencies except a narrow ``dip'' around the seed frequency, consistent with the phase-mismatch argument used to discard $\delta c_m$ in Eqs.~(\ref{eq:3-dp2}).\footnote{The term $\delta c_m$ is strongly phase-mismatched when $|\beta_{c,m}-\beta_{a,m}|L \gtrsim \pi$.  To first order, this depends on the group-velocity (or index) mismatch $\Delta\beta_1 = |\beta'(\omega_c) - \beta'(\omega_a)| \equiv \Delta n_g/c$ and the detuning: $\Delta f \gtrsim 1/(2\Delta\beta_1 L)$.  For example, for the parameters used in Fig.~\ref{fig:3-f8}, the narrow dip where Eq.~(\ref{eq:3-dp3}) fails to hold has a bandwidth of $\text{BW} = 2\Delta f = 1/(\Delta\beta_1 L) = 600$~GHz (about 5~nm), much narrower than the overall gain window.}

Fig.~\ref{fig:3-f8}(c) shows an example of mode-hopping in action: over the course of 200 round trips, the original CW seed (at 1.51~$\mu$m) is disrupted by amplified vacuum fluctuations, leading to a chaotic state; over long timescales ($N_{\rm rt} \gtrsim 10^5$), these states tend to stabilize to a CW oscillation at the mode with highest gain.  Interestingly, the MI gain spectrum is {\it not} maximized at the phase-matched wavelength, but is slightly offset at around 1.57~$\mu$m; the phase-mismatch of the seed mode causes the MI gain spectrum to be slightly offset from the undepleted gain spectrum.

\begin{figure*}[t!]
\begin{center}
\includegraphics[width=1.00\textwidth]{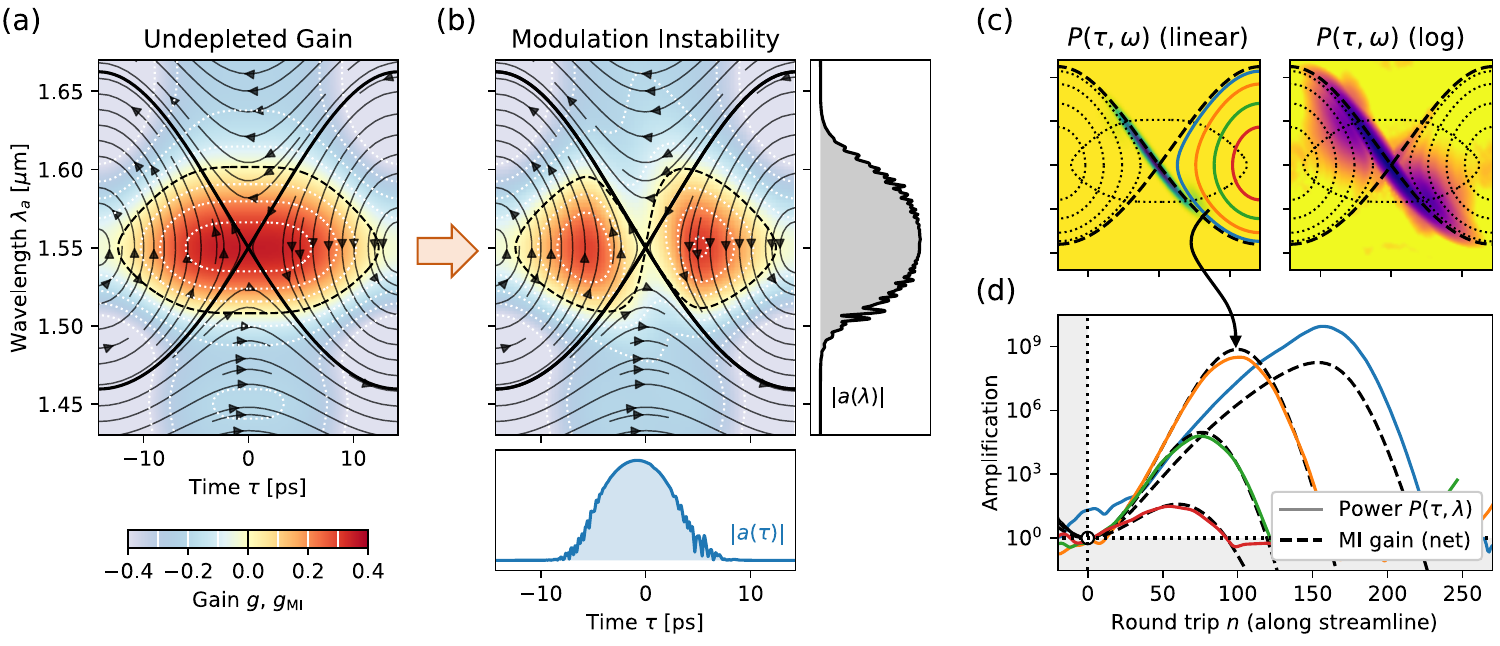}
\caption{Mode-hop instability mechanism.  (a) Phase portrait showing the phase-space dynamics $\dot{\tau}(t)$, $\dot{\omega(t)}$ and the OPA gain, which governs pulse formation.  (b) Once a pulse is formed, its stability to perturbations is governed by MI gain, and perturbations follow the same phase-space dynamics.  Here, we also plot the power $|a(\tau)|$ and spectrum $|a(\lambda)|$ of a representative FMCW pulse at the brink of mode-hopping instability.  (c) Spectrogram $P(\tau, \omega)$ plotted on a linear (left) and logarithmic (right) scale.  (d) Slices of $P(\tau, \omega)$ along four representative streamlines, showing close correspondence between the numerically computed fluctuations and the predictions of Eq.~(\ref{eq:3-intg}).  Parameters: $f_{\rm rep} = 35$~GHz, $\phi_p = 3$~rad, $\phi_a = 0.3$~rad.}
\label{fig:3-f9}
\end{center}
\end{figure*}

Next, we need to connect the theory of CW modulation instability to the phase-space dynamics of Sec.~\ref{sec:2-fmcw} to understand mode-hopping in QAM-OPO pulses.  Recall that comb-forming is a competition between gain $\dot{A} = \tfrac12(g(A, \omega) - \alpha(\tau)) A$, dispersion $\dot{\tau} = \delta'(\omega)/T$, and chirp $\dot{\omega} = \phi'(\tau)/T$, which trace out streamlines on ($\tau, \omega$) phase space, leading to a stable FMCW pulse along the separatrix of the phase portrait (Fig.~\ref{fig:3-f9}(a)).  Similarly, once this pulse is formed, we can study the dynamics of the perturbations $\delta a_m$ using the same phase portrait, but with the amplitude governed by MI gain $\delta \dot{A} = \tfrac12 (g_{\rm MI}(\tau,\omega) - \alpha(\tau)) \delta A$ instead (Fig.~\ref{fig:3-f9}(b)).

Note that, consistent with Fig.~\ref{fig:3-f8}(b), the MI gain is maximized near the phase-matched wavelength (here 1.55~$\mu$m) at the tails of the FMCW pulse, where the oscillating signal is detuned from phase-matching and pump depletion is weak.  However, these gain regions (red lobes in Fig.~\ref{fig:3-f9}(b)) are centered {\it away} from any dynamical fixed points, so the phase-space dynamics sweep away any perturbation before it has time to build up.  This is the key effect that allows the QAM-OPO to sustain broad combs that push the limits of the phase-matching envelope without being destabilized by mode-hopping.  As we see from the plots of the intensity $|a(\tau)|$ and spectrum $|a(\lambda)|$, the pulse in Fig.~\ref{fig:3-f9}(b) is on the verge of instability due to mode hopping, with amplified vacuum fluctuations already starting to disturb its tails.  This amplification is observed in Fig.~\ref{fig:3-f9}(c), where we plot the spectrogram on a linear scale (left) and log scale (right).  The log-scale plot shows clear exponential gain as we traverse streamlines of the phase-space dynamics, starting from vacuum fluctuations (yellow) and growing to be comparable to the background pulse amplitude.  Likewise, after leaving the gain region, the fluctuations decay back to vacuum due to negative net MI gain.

The most important metric governing mode-hop instability is the integrated gain
\beq
	G_{\rm int} = \int{\bigl(g_{\rm MI}(\omega, \tau) - \alpha(\tau)\bigr) \d n} \label{eq:3-intg}
\eeq
where the integral in Eq.~(\ref{eq:3-intg}) is taken over any streamline $(\tau(t), \omega(t))$, starting from the point where it enters the MI gain region.  As Fig.~\ref{fig:3-f9}(d) shows tracing four representative streamlines, $G_{\rm int}$ is fairly accurate at capturing the amplification of vacuum noise (as well as subsequent attenuation outside the gain region).  Net gain $G_{\rm int} > 0$ by itself does not destabilize the pulse, but there are two case where it can be disruptive:
\begin{enumerate}
	\item If gain is large enough that vacuum fluctuations become comparable to the pulse amplitude.  This depends on the OPO threshold but is roughly $G_{\rm int} \gtrsim 20$--$30$.
	\item If net gain $G_{\rm loop} = T^{-1} \oint{\bigl(g_{\rm MI}(\omega, \tau) - \alpha(\tau)) \d t}$, integrated over the {\it entire streamline} is positive.  Since the streamlines form loops, net gain over a loop will lead to runaway amplification of a vacuum signal.
\end{enumerate}

Armed with Eq.~(\ref{eq:3-intg}), we see that there are two factors that can contribute to the onset of mode-hop instability: if the MI gain $g_{\rm MI}(\omega,\tau)$ is too large, or if the ``dwell time'' in the gain region, i.e.\ the number of round trips when traversing a streamline, is too long.  Both factors are influenced by a number of experimental parameters, and studying how these parameters lead to mode-hopping will help us devise strategies to prevent it.

\begin{figure*}[t!]
\begin{center}
\includegraphics[width=1.00\textwidth]{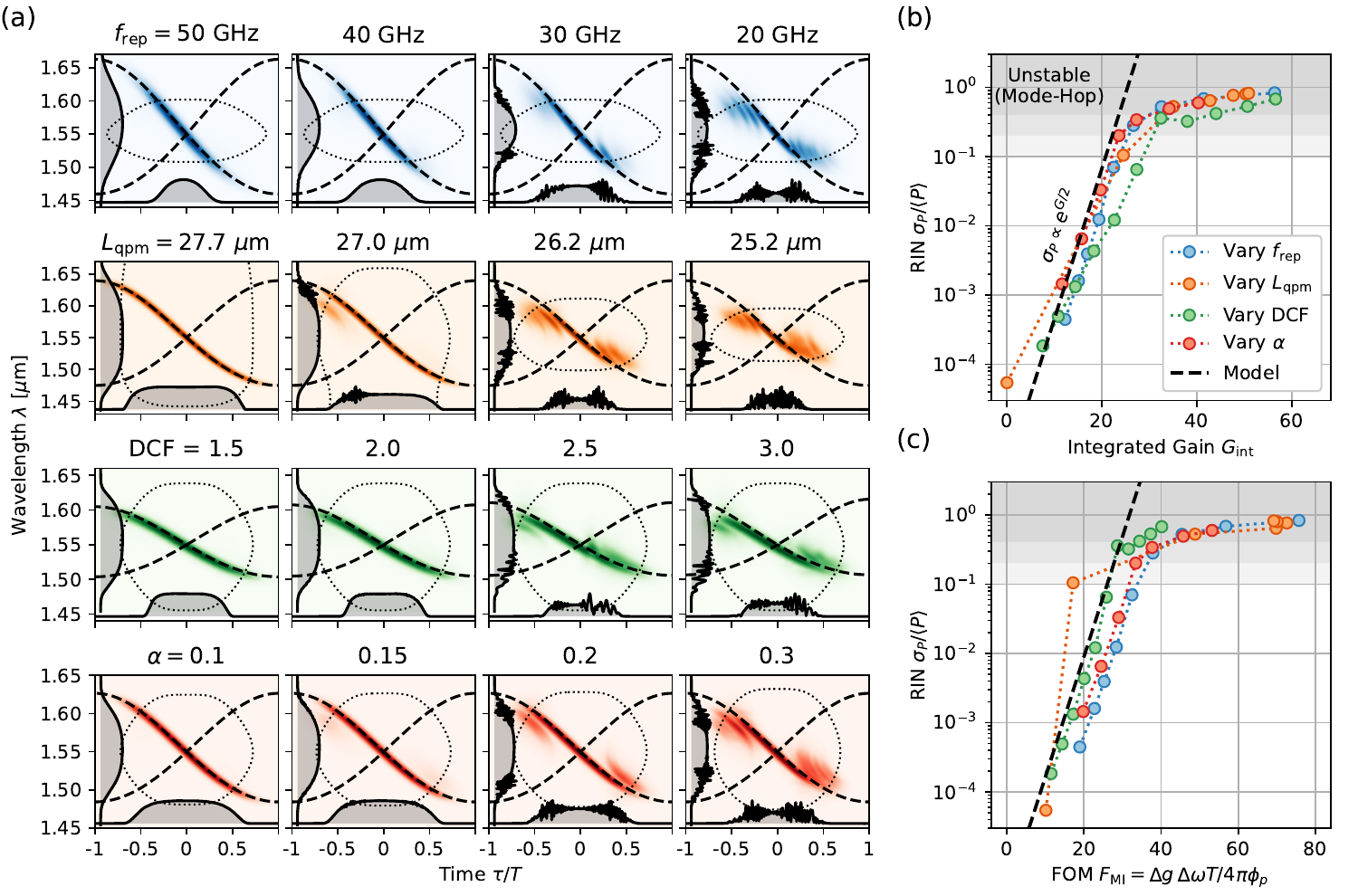}
\caption{(a) Onset of modulation instability when a parameter is varied.  Top to bottom: repetition rate $f_{\rm rep}$, QPM period (which controls signal-idler GVM and thus phase-matching bandwidth), dispersion compensation, and round-trip loss.  (b) Comb-line RIN as a function of integrated gain $G_{\rm int}$ for the highest-gain streamline.  (c) RIN as a function of MI figure of merit $F_{\rm MI}$, showing that mode-hopping can be avoided when $F_{\rm MI} \leq 25$.}
\label{fig:3-f10}
\end{center}
\end{figure*}

Fig.~\ref{fig:3-f10}(a) shows the onset of mode-hopping as a function of four parameters in the QAM-OPO.  From top to bottom, these are: (i) EOM frequency $f_{\rm rep}$, (ii) gain bandwidth (via adjusting $L_{\rm qpm}$), (iii) round-trip dispersion, and (iv) round-trip loss.  As each parameter is varied, the smooth FMCW spectrogram shows the growth of modes away from the separatrix at the tails of the pulse, indicative of mode-hopping to the higher-gain center frequencies; this in turn leads to chaotic interference patters in both the temporal amplitude $|a(\tau)|$ and spectrum $|a(\lambda)|$ (shown at the bottom and left of each subplot):

\begin{enumerate}
	\setcounter{enumi}{0} 
	\renewcommand{\theenumi}{\roman{enumi}} 
	\item The EOM frequency is directly related to the round-trip frequency chirp $\phi'(t) = \phi_p \Omega \sin(\Omega t) = 2\pi \phi_p f_{\rm rep} \sin(2\pi f_{\rm rep} t)$.  The number of round-trips needed to traverse the gain region will be approximately $\Delta\omega / f_{\rm rep}$, so the integrated gain scales as $G_{\rm int} \propto 1/f_{\rm rep}$.  Therefore, mode-hopping will be especially relevant when generating frequency combs of narrow spacing, where $f_{\rm rep}$ is small.
	\item The OPA gain bandwidth affects mode hopping, since MI only occurs when the OPO oscillates away from the frequency with maximum gain.  By varying the QPM period of the $\chi^{(2)}$ crystal, we can adjust the signal-idler group-velocity mismatch and thus shrink the gain bandwidth.  The second row of Fig.~\ref{fig:3-f10}(a) shows QAM-OPO for four separate crystals, showing how shrinking the gain window leads to mode-hopping even when all other parameters are equal.
	\item Dispersion compensation is a great way reduce RF power without sacrificing bandwidth, but it can also trigger mode-hopping (in addition to the loopback instability, Sec.~\ref{sec:3-loop}).  In the third row of Fig.~\ref{fig:3-f10}(a), we reduce $\beta_2 \rightarrow \beta_2/\text{DCF}$ by a constant factor by adding a fixed length of SMF28 fiber to the cavity; at the same time, to maintain the comb bandwidth, we reduce the PM phase $\phi_p \rightarrow \phi_p/\text{DCF}$.  Dispersion compensation (at fixed bandwidth) reduces the chirp rate by the same factor, correspondingly increasing $G_{\rm MI} \propto \text{DCF}$.
	\item Finally, high round-trip loss can lead to increased mode-hopping because gain and loss are always of the same order of magnitude: $g_{\rm MI} \sim g \sim \alpha$.  On the other hand, the phase-space dynamics that set the gain-region dwell time (chirp and dispersion) are independent of loss.  Therefore, we expect $G_{\rm int} \propto \alpha$, and increasing loss can destabilize the comb, as shown in the lower column of Fig.~\ref{fig:3-f10}(a).
\end{enumerate}

\begin{figure*}[t!]
\begin{center}
\includegraphics[width=1.00\textwidth]{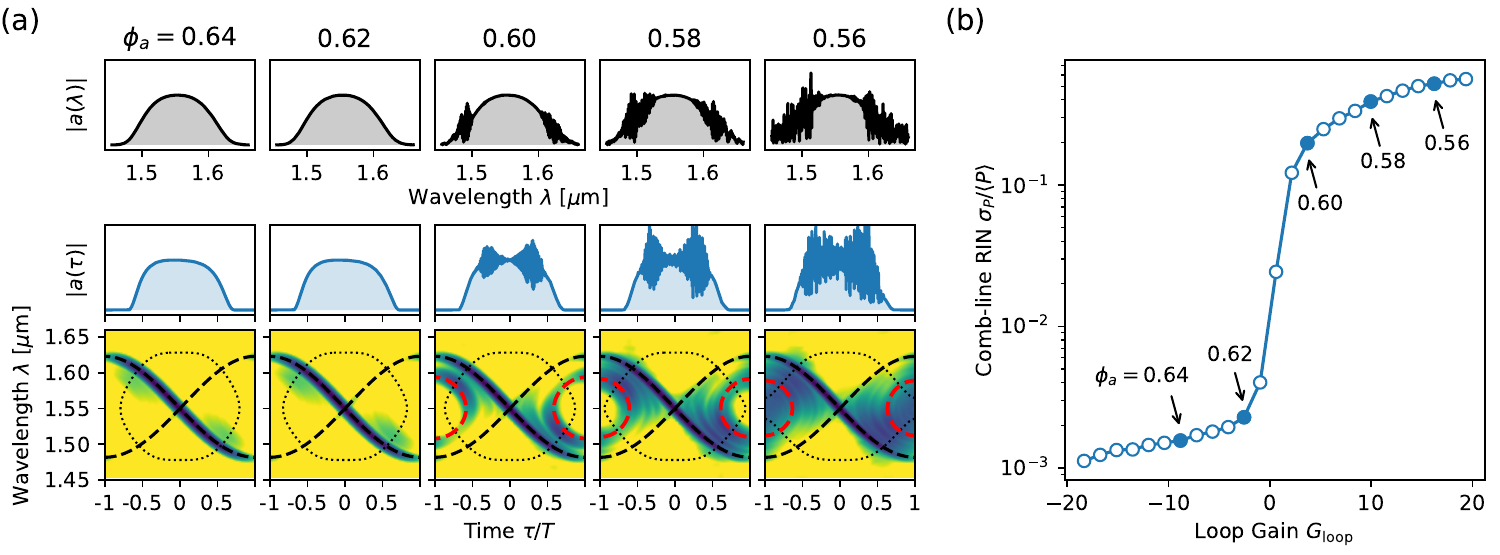}
\caption{Loop-gain instability.  (a) Reducing the IM phase $\phi_a$ leads to the sudden onset of fluctuations.  The spectrograms reveal these fluctuations center around the loop streamline with maximum gain (dashed red line).  (b) Comb-line RIN as a function of loop gain.  Parameters: $f_{\rm rep} = 25$~GHz, $\phi_p = 3$~rad, DCF = 0.5.}
\label{fig:3-f11}
\end{center}
\end{figure*}

To make this discussion more quantitative, we can calculate the comb-line relative intensity noise (RIN) $\sigma_P/\langle P\rangle$, which should increase to $O(1)$ when the comb becomes unstable.  Fig.~\ref{fig:3-f10}(b) plots the RIN as a function of the integrated gain $G_{\rm int}$ (Eq.~(\ref{eq:3-intg})) calculated for the four cases discussed above.  Vacuum fluctuations, which grow to an amplitude $\delta a \propto e^{G_{\rm int}/2}$ (recall $G_{\rm int}$ is power gain), interfere with the FMCW background to create a RIN that scales as $\sigma_P/\langle P\rangle \propto e^{G_{\rm int}/2}$.  This is consistent across all four scenarios, despite the very different mechanisms used to trigger mode-hop instability.

Using the pendulum model of Sec.~\ref{sec:2-pend}, we can derive a surprisingly good heuristic formula for $G_{\rm int}$, and consequently the onset of mode-hopping.  Recall that the dynamical timescale was found to be $\xi_n = (\Omega\sqrt{\delta_2\phi_p})^{-1} = \Delta\omega_{\rm max} T/8\pi\phi_p$ round trips (combining Eqs.~(\ref{eq:2-norm}) and (\ref{eq:2-dwmax})).  Assume that (i) the gain-region dwell time is comparable to this timescale, and (ii) the MI gain is approximately the difference $g_{\rm MI} \rightarrow \Delta g \equiv g_{\rm max} - g_{\rm min}$ between its maximum (at the phase-matched center frequency, $g(0)$) and minimum (at the antipodes, $g(\pm \Delta\Omega/2)$) values on the separatrix.  Eq.~(\ref{eq:3-intg}) then simplifies to:
\begin{align}
	G_{\rm int} & = \int{\bigl(g_{\rm MI}(\omega, \tau) - \alpha(\tau)\bigr) \d n} \nonumber \\
	& \approx (g_{\rm max}-g_{\rm min}) \xi_t 
	= \underbrace{\frac{\Delta g\, \Delta\omega_{\rm max} T}{8\pi \phi_p}
	}_{F_{\rm MI}} \label{eq:3-fmi}
\end{align}
We call this quantity $F_{\rm MI}$, the MI / mode-hopping figure of merit.  This figure is very easy to calculate, since $\Delta\omega_{\rm max}$ and $\Delta g$ both have analytic expressions (see Eqs.~(\ref{eq:2-gnorm}, \ref{eq:2-dwmax}), with $\Delta\beta_{\rm max} = \Delta\beta(\pm\Delta\omega_{\rm max}/2)$ below):
\begin{align}
	\Delta\omega_{\rm max} & = 4\sqrt{\phi_p/\delta_2} \nonumber \\
	\Delta g & = p^2\alpha\bigl[1 - \text{sinc}^2\bigl(\tfrac12\Delta\beta_{\rm max} L\bigr)\bigr] \nonumber \\
	& \quad \sim \begin{cases}
		p^2\alpha (\Delta\beta_{\rm max} L)^2/12 & (\Delta\beta_{\rm max} L \ll 1) \\
		p^2\alpha & (\Delta\beta_{\rm max} L \gtrsim 2\pi)
		\end{cases}
\end{align}
Fig.~\ref{fig:3-f10}(c) plots the comb-line RIN against $F_{\rm MI}$, where we consistently see the onset of instability when $F_{\rm MI} \gtrsim 20$--$30$.

Since the number of comb lines in the FMCW model is bounded by the separatrices $N_{\rm comb} < \Delta\omega_{\rm max} T/2\pi$, mode-hopping instability limits the number of comb lines in the QAM-OPO.  Requiring $F_{\rm MI} < 25$ based on what we empirically see from Fig.~\ref{fig:3-f10}(c), one finds:
\beq
	N_{\rm comb} = \frac{2 F_{\rm MI} \phi_p}{\Delta g} < \frac{100}{p^2 (1 - \text{sinc}^2(\Delta\beta_{\rm max} L/2))} \frac{\phi_p}{\alpha} \label{eq:3-ncomb}
\eeq
The $\phi_p/\alpha$ dependence should ring a bell: the 3-dB bandwidth of the EO comb is given by $N_{\rm eo} = 2 \log(2) \phi_p/\alpha$ (Eq.~(\ref{eq:2-neta}), see also \cite{zhang2019broadband}).  Alas, the number of comb lines for both frequency combs is limited RF power and cavity finesse, with a scaling $N \sim \phi_p/\alpha$.  However, the constant factor in Eq.~(\ref{eq:3-ncomb}) is large, so that even in the worst-case scenario where the tails of the comb are significantly phase-mismatched (and the sinc term in the denominator goes to zero), the QAM-OPO can still support at least $30\times$ more comb lines:
\begin{align}
	\frac{N_{\rm opo}}{N_{\rm eo}} & = \frac{100}{p^2 (1 - \text{sinc}^2(\Delta\beta_{\rm max} L/2))} \frac{\phi_p}{\alpha}
	\bigg{/} \frac{2\log(2)\phi_p}{\alpha} \nonumber \\
	& = \frac{50}{\log(2) p^2} \frac{1}{1 - \text{sinc}^2(\Delta\beta_{\rm max} L/2)} \gtrsim 30
\end{align}
where we have assumed $p \sim 1.57$, which is the pump power that maximizes conversion efficiency in a CW-OPO (see Appendix~\ref{sec:a-opo}).  If the phase-matching bandwidth is broad compared to the comb bandwidth, or if the QAM-OPO is pumped closer to threshold, it can support even more comb lines.

Finally, as mentioned earlier, the comb can experience an instability if there is net gain over any closed streamline in phase space, i.e.\ $G_{\rm loop} > 0$.  The result is analogous to wraparound instability, although it arises from closed characteristics rather than the wrapping around of the separatrix.  To illustrate an example, consider the OPO in Fig.~\ref{fig:3-f11}, where we slowly reduce the amplitude modulation $\phi_a$.  Here, at about $\phi_a = 0.3$, the net gain of the loop goes positive, leading to a chaotic structure centered around the fixed point at $\tau = \pm T/2$ (Fig.~\ref{fig:3-f11}(a)).  Note that, in this case, the onset of oscillations is sudden (Fig.~\ref{fig:3-f11}(b)) and occurs exactly when $G_{\rm loop} = 0$.  This sharp turn-on is in contrast to the gradual emergence of fluctuations in Fig.~\ref{fig:3-f10}, and indicates that the instability is a true bifurcation (as opposed to merely exponential amplification of vacuum noise).

\subsection{Conditions for Stable Comb Generation}

To conclude this section, Table~\ref{tab:t3} lists the three conditions that must roughly be satisfied in order to achieve stable comb generation in the QAM-OPO.  Each condition is associated with a separate instability, and has its own figure of merit.  For wraparound instability, it is the field attenuation over the loss region, $e^{-\gamma}$ (Eq.~(\ref{eq:3-egam})); for loopback instability, the phase relative to its TOD-limited maximum value $\phi_p/\phi_{\rm max}$ (Eq.~(\ref{eq:3-lpmax})); and for mode-hop instability, the integrated MI gain $F_{\rm MI}$ (Eq.~(\ref{eq:3-fmi})).

\begin{table}[htbp]
\begin{center}
\begin{tabular}{c|cl}
	\hline\hline
	{\bf Instability} & \multicolumn{2}{c}{\bf FoM and Condition}  \\ 
	\hline
	Wraparound & $\gamma = \frac{\sqrt{2} (p^2-1)\alpha_0}{6\pi f_{\rm rep}\sqrt{\delta_2\phi_p}} \Bigl[\frac{4\phi_a^2}{(p^2-1)\alpha_0} - 1\Bigr]^{3/2}$ 
		\!\!\!\!\!&\!\!\!\!\! $ > 3$ \\
	Loopback & $\vphantom{\Bigr|}\phi_p/\phi_{\rm max} = \frac{\phi_p \delta_2^3}{3\delta_3^2}$ &\!\!\!\!\! $ < 1$ \\
	Mode-Hop & $\vphantom{\Bigr|}F_{\rm MI} = \frac{\Delta g\, \Delta\omega_{\rm max}}{8\pi \phi_p f_{\rm rep}}$ &\!\!\!\!\! $ < 20$ \\
	\hline\hline
\end{tabular}
\caption{Wraparound, loopback, and mode-hop instability figures of merit and requirements for stable operation.}
\label{tab:t3}
\end{center}
\end{table}

\section{Dynamical Tuning} \label{sec:tuning}

A big advantage of OPOs is that they are broadband and highly tunable, as they rely on virtual transitions for gain, which typically have greater bandwidth than real transitions.  Indeed, the gain of an OPO is almost never set by the bandwidth of the intrinsic $\chi^{(2)}$ interaction, but by phase-matching, and with appropriate dispersion engineering, this phase-matched bandwidth can be large, even approaching or exceeding an octave \cite{jankowski2022quasi, sekine2023multi}.  This raises the question as to whether the QAM-OPO is also tunable.  In this section, we explore two ways to tune the QAM-OPO: varying the EOM RF frequency, which is fast but limited by fixed OPA phase-matching (Sec.~\ref{sec:4-rftune}), and varying the pump wavelength, which is slower but gives a wider tuning range (Sec.~\ref{sec:4-lamtune}).  Tuning is a repeatable, mode-hop-free process that enhances the flexibility of the QAM-OPO over conventional comb sources, whose center frequency is typically fixed by either the pump laser\cite{zhang2019broadband, herr2014temporal} or the laser gain maximum \cite{haus2000mode}.

\begin{figure*}[t!]
\begin{center}
\includegraphics[width=1.00\textwidth]{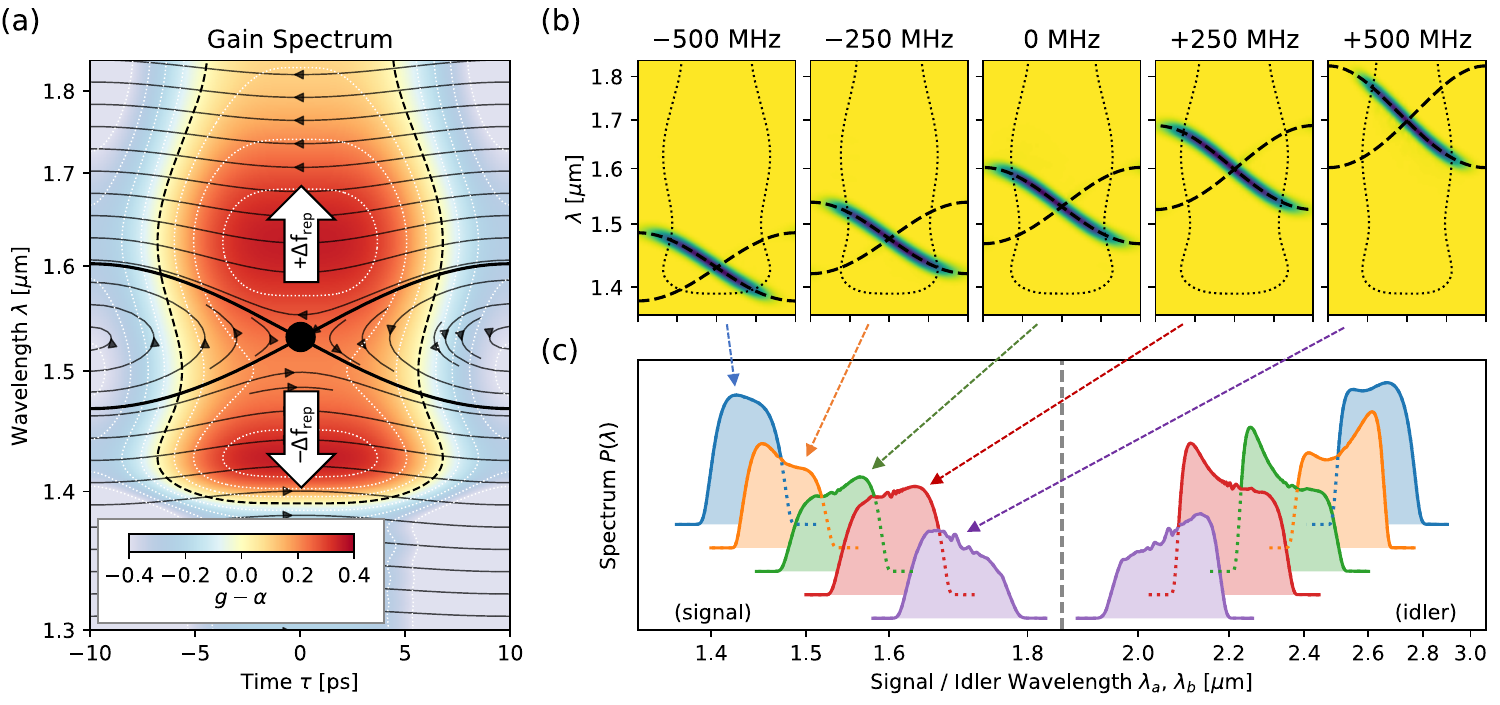}
\caption{RF tuning of the QAM-OPO.  (a) Gain spectrum and streamlines for a QAM-OPO optimized for broadband gain around $1.55~\mu$m ($\lambda_c = 929$~nm, $L_{\rm qpm} = 27.4~\mu$m, $f_{\rm rep} = 50$~GHz, $\phi_p = 1.5$, $\phi_a = 0.38$, see Appendix~\ref{sec:bw}).  Changing $f_{\rm rep}$ adds a group delay that shifts the saddle-point frequency.  (b) Spectrogram of the OPO output as $\Delta f_{\rm rep}$ is tuned from $-500$~MHz to $+500$~MHz.  The spectrum follows the separatrix, which shifts in frequency to following Eq.~(\ref{eq:4-dw}).  
(c) Signal and idler spectra (linear scale), showing full tuning of the signal and idler combs over the octave $\lambda \in [1.4, 2.8]~\mu$m (excluding degeneracy).}
\label{fig:4-f3}
\end{center}
\end{figure*}

\subsection{Tuning by RF Frequency}
\label{sec:4-rftune}

Recall that the QAM-OPO forms an FMCW pulse shaped by the phase-space dynamical equations $\d\tau/\d n = \delta'(\omega)$, $\d\omega/\d n = -\phi'(\tau)$, where $\delta(\omega)$ is the cavity round-trip dispersion and $\phi(\tau)$ is the EOM phase (Eqs.~(\ref{eq:2-char})); these lead to group delay and chirp, respectively.  The QAM-OPO pulse centers around the saddle point where $\delta'(\omega) = \phi'(\tau) = 0$, which we engineer to be near the maximum of the gain spectrum in order to suppress competing processes (Fig.~\ref{fig:4-f3}(a)).  Now let us detune the EOM's RF frequency $f_{\rm rep}$ relative to the cavity FSR.  This detuning $\Delta f_{\rm rep}$ is equivalent (up to some change in the carrier offset) to an added cavity group delay $\Delta T = \Delta f_{\rm rep}/f_{\rm rep}^2$, which adds a linear term to the cavity dispersion: $\delta(\omega) \rightarrow \delta(\omega) + \omega\,\Delta T$.  Expanding $\delta(\omega)$ to second order, we find that the frequency center shifts by:
\beq
	\Delta\omega = -\Delta T / \delta_2 \label{eq:4-dw}
\eeq
Therefore, increasing the RF frequency redshifts the comb, while decreasing the frequency blueshifts it.  To first order, this relation is linear, and the RF-to-optical frequency-shift multiplier $\Delta\omega/\Delta\Omega$ (here $\Omega = 2\pi f_{\rm rep}$) is:
\beq
	\frac{\Delta\omega}{\Delta\Omega} = -\frac{2\pi}{\delta_2 \Omega^2} = -\frac{1}{2 \pi \delta_2 f_{\rm rep}^2} \label{eq:4-dwdw}
\eeq
The system studied in Fig.~\ref{fig:4-f3} is designed to have particularly high gain bandwidth, using the techniques of group-velocity matching and intentional phase-mismatch discussed in Appendix~\ref{sec:bw}.  This leads to an OPO whose gain window, at an EOM drive of $\phi_p = 1.5$, is much broader than the actual frequency comb produced, i.e.\ a system that could make use of frequency tuning.  It has $\delta_2 = 2000$~fs$^2$ (Table~\ref{tab:t1} parameters) and $f_{\rm rep} = 50~\text{GHz}$, so Eq.~(\ref{eq:4-dwdw}) gives a multiplier of $30000\times$.  Having a large multiplier is convenient, as even small shifts in the EOM frequency lead to significant shifts in the comb center frequency, although an excessively large multiplier may make the comb overly sensitive.  We can recast Eq.~(\ref{eq:4-dwdw}) into a more intuitive form by replacing the GVD with the comb bandwidth and EOM phase, through the pendulum-model formula $\Delta \omega_{\rm BW} = 4\sqrt{\phi_p/\delta_2}$.  The result depends on the phase and the number of comb lines:
\beq
	\frac{\Delta\omega}{\Delta\Omega} = \frac{\pi}{8 \phi_p} \Bigl(\frac{\Delta\omega_{\rm BW}}{\Omega}\Bigr)^2
	= \frac{\pi N_{\rm comb}^2}{8\phi_p} \label{eq:4-dwdw2}
\eeq
In Fig.~\ref{fig:4-f3}(b-c), we vary the RF detuning to tune the comb.  The spectrograms are plotted in Fig.~\ref{fig:4-f3}(b).  Here, an RF tuning range of 1~GHz (2\% of $f_{\rm rep}$) is sufficient to sweep the signal frequency from 1.4~$\mu$m to just below the degeneracy point (around 1.8~$\mu$m).  When considering both signal and idler spectra (Fig.~\ref{fig:4-f3}(c)), this provides for a full tuning range of just over an octave.  Of course, getting an octave of tuning range is unremarkable for an OPO; what makes this unique is that we are tuning a {\it frequency comb}, not just a CW signal, over such a wide range.

\begin{figure*}[t!]
\begin{center}
\includegraphics[width=1.00\textwidth]{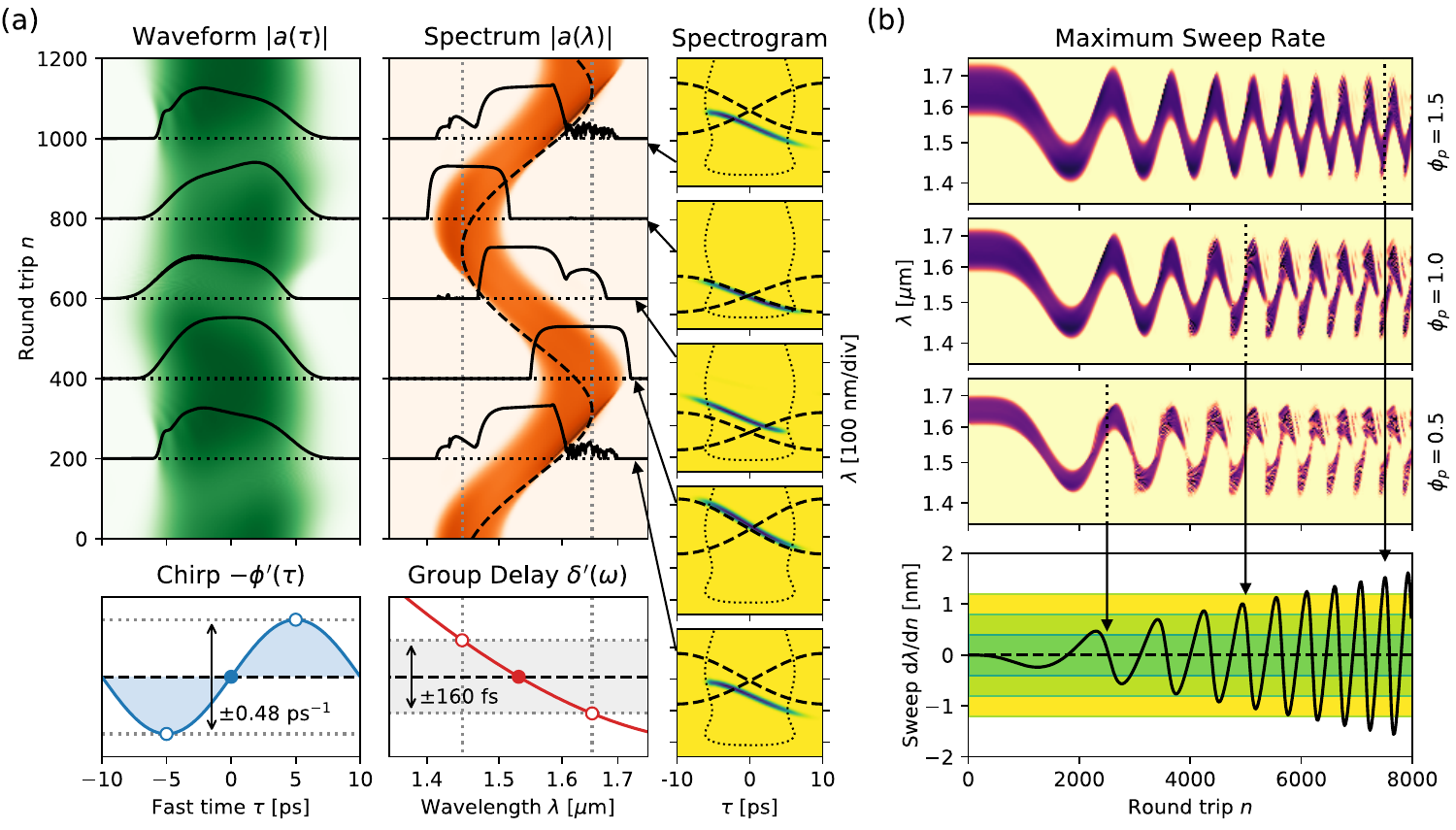}
\caption{Rapid RF wavelength tuning of the QAM-OPO from Fig.~\ref{fig:4-f3}.  (a) Waveform and spectrum of the OPO signal output when the RF frequency is tuned over a range of $f_{\rm rep}\pm0.4$~GHz, which corresponds to an effective group delay tuning of $\pm 160$~fs, with a period of 1000 round-trips.  (b) Spectra $|a(\lambda)|$ for three EO-OPOs with different PM modulation amplitudes $\phi_a = 1.5$, 1.0, and 0.5.  The tuning rate is linearly ramped, showing the onset of instabilities when the frequency sweep is too fast.  The bottom curve shows the sweep rate and bounds $\pm 4\pi f_{\rm rep} \phi_p$ ($2\times$ the result of Eq.~(\ref{eq:4-dwdtmax})) for the three $\phi_a$ values, showing that the maximum sweep rate scales linearly with $\phi_a$.}
\label{fig:4-f4}
\end{center}
\end{figure*}

Another advantage of RF tuning is that it can be {\it fast}, since it doesn't rely on any slow thermal dynamics or laser wavelength tuning, only on the frequency of the RF source.  But even with a perfectly frequency-agile RF source, how fast can comb tuning occur?  Fig.~\ref{fig:4-f4}(a) gives and example of rapid QAM-OPO tuning using the same RF-tuned OPO.  Here, we vary $f_{\rm rep}$ by $\pm 0.4$~GHz (which corresponds to a group delay variation of $\pm 160$~fs), which through the relation $\delta'(\omega) = \Delta T$ leads to a center frequency tuning range of $\lambda \in [1.45~\mu{\rm m}, 1.65~\mu{\rm m}]$.

Note the dynamical lag in frequency tuning: While the pulse spectrum $|a(\lambda)|$ largely follows the predicted steady-state tuning curve, there is a delay of about 100 round-trips (2~ns at 50~GHz).  In addition to this lag in frequency response, the spectrograms show that the pulse is shifted in time relative to the center.  Consulting the waveform $|a(\tau)|$ plots, it appears that the pulse is shifting left ($\tau < 0$) when the ECPO is redshifting, and right ($\tau > 0$) when blueshifting.  This would be consistent with the chirp $\d\omega/\d n = -\phi'(\tau)$, plotted on the bottom, which is positive for $\tau > 0$ and negative for $\tau < 0$.  When redshifting, the pulse shifts to the left and experiences an average negative EOM chirp, while the opposite happens when blueshifting.

In fact, the EOM chirp
\beq
	\frac{\d\omega}{\d n} = -\phi'(\tau) = 2\pi f_{\rm rep} \phi_p \sin(\Omega\tau)
\eeq
%
gives an approximate bound on the maximum tuning sweep rate of the QAM-OPO.  At a high level, if we expect the chirp to be the dominant mechanism shifting the OPO's center frequency (in reality it is a more complex nonlinear process involving gain and loss), then the maximum sweep rate should be:
\beq
	\frac{\d\omega}{\d n}\Bigr|_{\rm max} = 2\pi f_{\rm rep} \phi_p \label{eq:4-dwdtmax}
\eeq
In simulations, it appears we can go a bit faster, but Eq.~(\ref{eq:4-dwdtmax}) gives a good understanding of the dependence of sweep rate on OPO parameters.  In Fig.~\ref{fig:4-f4}(b), we slowly ramp the frequency sweep rate for three ECPOs with different PM phase shifts $\phi_p = 1.5$, $\phi_p = 1.0$, and $\phi_p = 0.5$.  Predictably, when the wavelength sweep is too fast, the comb dynamics cannot keep up, and the pulse goes unstable.  This instability sets in at different sweep rates, and the scaling follows the dependence $(\d{\omega}/\d n)_{\rm max} \propto \phi_p$ expected from Eq.~(\ref{eq:4-dwdtmax}).  However, the actual OPO can tolerate about $2\times$ faster wavelength sweeps than the equation would predict: for the $\phi_p = 1.5$ case as an example, with $f_{\rm rep} = 50$~GHz as before, Eq.~(\ref{eq:4-dwdtmax}) predicts a maximum sweep rate of $\dot{\omega}_{\rm max}/2\pi = 75$~GHz/round-trip (0.6~nm/round-trip), whereas the QAM-OPO appears to tolerate up to 1.2~nm/round-trip (yellow region in Fig.~\ref{fig:4-f4}(b)).  The discrepancy is likely due to the other nonlinear dynamics, e.g.\ gain and pump depletion, that have not been considered in the simple chirp-based frequency sweep model leading to Eq.~(\ref{eq:4-dwdtmax}).

\subsection{Tuning by Pump Wavelength, Temperature}
\label{sec:4-lamtune}

While RF tuning is fast, it will always be limited by the phase-matching bandwidth of the OPA crystal.  This is not a big problem as long as one operates close enough to the gain medium's zero-dispersion wavelength, where the bandwidth is very large (as described in Appendix~\ref{sec:bw}), but far from that wavelength, there will be relatively limited tuning range.  For example, with a bulk LiNbO$_3$ gain medium, it is difficult to find conditions (at a fixed pump wavelength) where the QAM-OPO is continuously tunable through all the major telecom bands (O-, E-, S-, C-, L-, and U-bands, 1260-1675~nm), due to phase-matching constraints.

\begin{figure*}[t!]
\begin{center}
\includegraphics[width=1.00\textwidth]{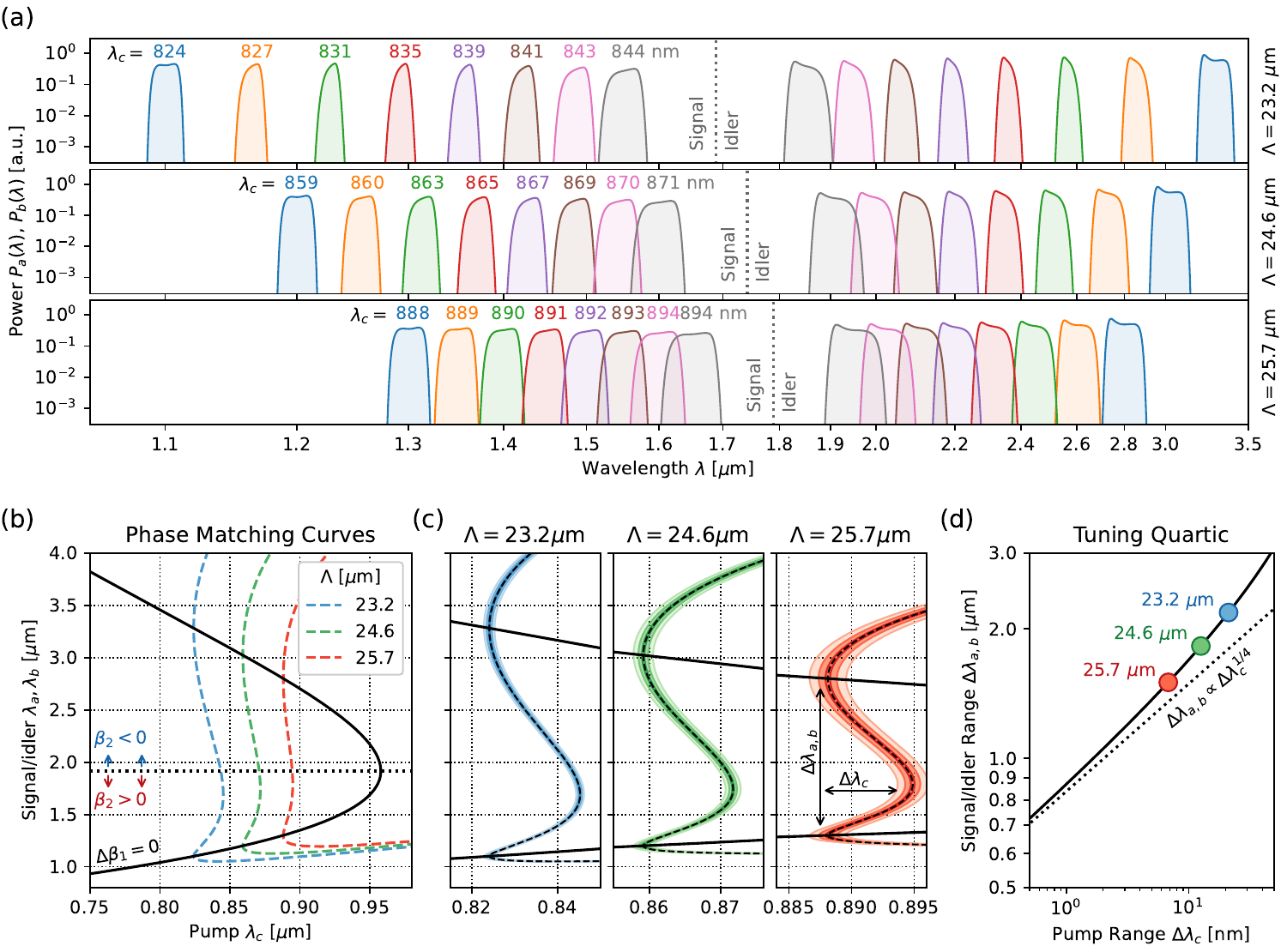}
\caption{Pump-wavelength tuning of the QAM-OPO.  (a) Signal and idler comb spectra for three QAM-OPOs with different QPM lengths.  Parameters: $\phi_p = \phi_a = 0.4$, $f_{\rm rep} = 100$~GHz.  QPM lengths, top to bottom: $L_{\rm qpm} = 23.2~\mu$m, $24.6~\mu$m, $25.7~\mu$m.  (b-c) Phase-matching curves for these OPOs.  (d) Relation between pump and signal/idler tuning range, following the quartic $\Delta\lambda_{c} \propto \Delta\lambda_{a,b}^4$ (Eq.~(\ref{eq:4-lquart})).}
\label{fig:4-f5}
\end{center}
\end{figure*}

However, by supplementing RF tuning with pump wavelength tuning, we can reach all these bands.  While standard RD tuning constrains us to fixed phase-matching window (fixed pump $\lambda_c$), by varying $\lambda_c$, we can explore phase-matching over a much wider range of conditions.  Practically speaking, this allows reasonably broadband combs all the way from the GV-matching wavelength (where $\beta_1(\lambda_a) = \beta_1(\lambda_b)$) to the degeneracy point $\lambda_a = \lambda_b = 2\lambda_c$, with relatively limited tuning of the pump laser.  For a fixed-wavelength pump, the same effect can also be achieved by temperature tuning the crystal, which shifts the phase-matching curve due to the wavelength-dependent thermo-refractive coefficient ${\d}n(\lambda)/{\d}T$ \cite{gayer2008temperature}.  Note that pump- and temperature-tuning do not replace RF-tuning in the QAM-OPO; one still needs to move the saddle point by adjusting $\Delta f_{\rm rep}$ (either through the RF frequency or cavity FSR, Fig.~\ref{fig:4-f3}(a)) to keep the saddle within the gain region and maintain a stable comb.  However, the wavelength range increases over standard RF tuning.

\newcolumntype{C}{>{\centering\arraybackslash}p{1.6cm}}
\newcommand{\GRY}{\cellcolor{black!10}}
\newcommand{\GRN}{\cellcolor{green!25}}
\newcommand{\GRNN}{\cellcolor{green!35}}
\newcommand{\YEL}{\cellcolor{yellow!40}}
\newcommand{\RED}{\cellcolor{red!25}}
\newcommand{\BLU}{\cellcolor{gray!20}}

\begin{table*}[htbp]
\begin{center}
\begin{tabular}{c|CC|CC|C|CC}
\hline\hline
& \multicolumn{2}{c|}{\bf DKS} & \multicolumn{2}{c|}{\bf EO Comb} & {\bf MLL} & \multicolumn{2}{c}{\bf OPO Comb} \\
& Bright & Dark & EO & RE-EO & & FM & \GRY {\it QAM} \\ \hline
Efficiency & \RED Low & \YEL Mid & \GRN $O(1)$ & \RED Low & \GRN $O(1)$ & \GRN $O(1)$ & \GRNN $O${\it (1)} \\
Bandwidth & \GRN Good & \YEL OK & \RED Limited & \YEL OK & \RED Limited & \GRN Good & \GRNN {\it Good}  \\
$\lambda$ tuning & \multicolumn{2}{c|}{\RED Locked to Pump} & \multicolumn{2}{c|}{\RED Locked to Pump} & \YEL Limited & \YEL Pump & \GRNN {\it RF-Tuned} \\
Spectral Flatness & \RED sech & \RED sech-like & \YEL Bessel & \RED sech & \RED sech & \YEL Bessel & \GRNN {\it Flat} \\
\hline
RF power & \multicolumn{2}{c|}{\GRN None} & \RED High & \YEL Mid & \YEL Varies & \GRN Low & \GRNN {\it Low} \\
Turn-on & \multicolumn{2}{c|}{\RED Sensitive} & \multicolumn{2}{c|}{\GRN Turn-key} 
& \YEL Varies & \GRN Turn-key & \GRNN {\it Turn-key} 
\\ 
Cavity locking & \RED Yes & \RED Yes & \GRN No & \RED Yes & \GRN No & \RED Yes (DR) & \GRNN {\it No (SR)} \\ 
\hline
& \cite{herr2014temporal} & \cite{xue2015mode} & \cite{torres2014optical} & \cite{zhang2019broadband} & \cite{haus2000mode} & \cite{stokowski2024integrated} & \GRY {\it this work} \\
\hline\hline
\end{tabular}
\caption{Comparison of the QAM-OPO and leading optical frequency comb platforms.  DKS: dissipative Kerr soliton; MLL: mode-locked laser; RE-EO: resonator-enhanced electro-optic; DR: doubly-resonant; SR: singly-resonant.}
\label{tab:7-t1}
\end{center}
\end{table*}

Fig.~\ref{fig:4-f5}(a) shows comb formation in a QAM-OPO under pump-wavelength tuning for three QPM periods: 23.2~$\mu$m, 24.6~$\mu$m, and 25.7~$\mu$m, showing tuning ranges significantly wider than what we achieved in Fig.~\ref{fig:4-f3}, especially for shorter wavelengths.  In Fig.~\ref{fig:4-f5}(b-c), we show the phase-matching curves for these three QPM periods.  Let us take the full tuning range to be between the group-velocity matching ($\Delta\beta_1 = 0$, black in figure) and degeneracy; technically, one can tune outside this range as well, but the bandwidth of the comb falls off rapidly due to dispersion.  We can derive a formula relating the pump tuning range $\Delta\lambda_c$ to the signal/idler tuning range $\Delta\lambda_{a,b}$ (shown in  Fig.~\ref{fig:4-f5}(c)) by Taylor-expanding the phase mismatch $\Delta\beta$ about the degeneracy point $\omega + \omega \leftrightarrow 2\omega$, namely $\omega_c = 2\omega + \delta\omega_c$, $\omega_{a,b} = \omega_c/2 \pm \delta\omega_a$, and assuming phase-matching at degeneracy:
\begin{align}
	\Delta\beta(\delta\omega_c, \delta\omega_a) & = 
		\beta(2\omega + \delta\omega_c) -
		\beta(\omega + \tfrac12\delta\omega_c + \delta\omega_a) \nonumber \\
		& \quad - \beta(\omega + \tfrac12\delta\omega_c - \delta\omega_a) - 2\pi/L_{\rm qpm} \nonumber \\
	& = (\beta_1(2\omega) - \beta_1(\omega)) \delta\omega_c - \beta_2(\omega) \delta\omega_a^2 \nonumber \\
		& \quad - \frac{\beta_4(\omega)}{12} \delta\omega_a^4 + O(\delta\omega_c^2,\,\delta\omega_a^2\delta\omega_c,\,\delta\omega_a^6)
\end{align}
The phase-matching curves (dashed in Fig.~\ref{fig:4-f5}(b-c)) correspond to $\Delta\beta = 0$; solving for the pump shift $\delta\omega_c$, we find to leading order in the Taylor series,
\beq
	\delta\omega_c = \frac{1}{\beta_1(2\omega)-\beta_1(\omega)} \Bigl[\beta_2(\omega)\delta\omega_a^2 + \frac{\beta_4(\omega)}{12} \delta\omega_a^4\Bigr]
\eeq
This is a quartic, and $\beta_2$ and $\beta_4$ have opposite signs in the curves studied in Fig.~\ref{fig:4-f5} (when they have the same sign, tuning is significantly less broadband away from degeneracy).  As a result, there is a maximum pump frequency shift $\delta\omega_c$, corresponding to the signal/idler group-velocity matched case.  This maximum occurs when:
\begin{subequations}\begin{align}
	\delta\omega_a & = \sqrt{-6\beta_{2}(\omega)/\beta_4(\omega)} \\ 
	\delta\omega_c & = \frac{-3(\beta_2(\omega))^2}{(\beta_1(2\omega)-\beta_1(\omega)) \beta_4(\omega)}
	= \frac{-\beta_4(\omega)}{12(\beta_1(2\omega)-\beta_1(\omega))} \delta\omega_a^4
\end{align}\end{subequations}
Converting this to wavelength perturbations via $\delta\omega = -(2\pi c/\lambda^2)\delta\lambda$, and noting that $\Delta\lambda_{a,b}$ is defined as {\it twice} the signal perturbation (since the range is $[-\delta\omega_a, +\delta\omega_a]$, covering both signal and idler), we get:
\beq
	\Delta\lambda_c = \frac{\pi^3 c^3 \beta_4(\omega)}{96\lambda_a^6 (\beta_1(2\omega)-\beta_1(\omega))} \Delta\lambda_{a,b}^4 \label{eq:4-lquart}
\eeq
This evaluates to 2~nm/$\mu$m$^4$ for a signal at the zero-dispersion wavelength $\omega = \omega_{\rm zdw}$, where the approximation is most appropriate.  In Fig.~\ref{fig:4-f5}(d), we plot the actual $\Delta\lambda_{a,b}$ vs.\ $\Delta\lambda_c$ curve against this quartic, showing good agreement (20-30\% error in $\Delta\lambda_{a,b}$) despite the coarse assumptions made for such broadband tuning.

\section{Conclusion} \label{sec:conc}

This paper has proposed a new type of frequency comb source, the QAM-OPO, based on hybrid AM/FM mode-locking of a parametric oscillator.  The QAM-OPO allows for stable, broadband, high-efficiency, turnkey comb generation in an OPO with moderate RF power.  The comb spectrum displays a high degree of flatness over the whole bandwidth, with a steady-state waveform that can be accurately described using an FMCW pulse ansatz.  We numerically simulated the QAM-OPO in a wide range of conditions and found that the dynamics are accurately understood in terms of a phase-space model, where comb formation is a balancing act involving gain and loss, amplitude modulation, chirp, and dispersion.  This picture is profoundly different from the soliton picture governing Kerr \cite{herr2014temporal, xue2015mode} or MLL combs \cite{haus2000mode}, as the waveform is chirped and quasi-CW, simultaneously broadband and delocalized in time.  The comb chirp can be cancelled with dispersion compensation, leading to a near transform-limited pulse.  Comb formation is stable under a wide range of conditions, but the phase-space model provides us with quantitative insights into the instabilities that can occur, whose onset can be quantified using three stability figures of merit $\gamma$, $\phi_p/\phi_{\rm max}$, and $F_{\rm MI}$.  The comb is rapidly tunable by adjusting the repetition-rate mismatch $\Delta f_{\rm rep}$ between the RF drive and cavity FSR, and the tuning range, set by the gain bandwidth of the OPA crystal, and be increased by also wavelength-tuning the pump, consistent with the well-known behavior of CW OPOs \cite{harris1969tunable, eckardt1991optical}.

Since its inception, the optical frequency comb has evolved as the underlying photonic technology progresses.  Originally restricted to tabletop or fiber-optic laser setups, the field saw a revolution with the introduction of compact, high-$Q$ microresonators \cite{savchenkov2004kilohertz, moss2013new}, enabling a new generation of broadband, low-power, compact dissipative Kerr soliton (DKS) combs.  Similarly, emerging integrated $\chi^{(2)}$ platforms (most notably LiNbO$_3$ \cite{zhu2021integrated} but also GaAs \cite{chang2018heterogeneously} and nitrides \cite{jena2019new, yoshioka2021strongly, li2024demonstration}) will enable entirely new integrated comb sources, such as OPO combs \cite{stokowski2024integrated}.  In this context, it is useful to compare the QAM-OPO to existing comb sources, both based on $\chi^{(3)}$ and $\chi^{(2)}$ effects, as well as EOM and laser mode-locking.  Table~\ref{tab:7-t1} gives a high-level comparison of the QAM-OPO and seven popular comb types, with respect to seven important properties for comb generation.  The frequency comb literature is very diverse so this table is far from exhaustive; e.g.\ it oversimplifies the diverse MLL physics (passive mode-locking, FM mode-locking, hybrid mode-locking \cite{chang2022integrated}) into a single category, omits many less well-known schemes (Pockels comb \cite{bruch2021pockels}, parametrically-driven DKS \cite{englebert2021parametrically}), and does not include a number of other important comb properties (FSR, phase noise).  But it provides enough information to give a general sense of the state of the field and cases where the QAM-OPO offers an advantage:

\begin{itemize}
	\item {\it Efficiency.}  Both bright DKS \cite{coen2013universal} and resonator-enhanced EO (RE-EO) \cite{zhang2019broadband} combs suffer from an efficiency-bandwidth tradeoff $\eta \propto 1/N_{\rm comb}$, and broadband combs usually have efficiencies of at most a few percent.  Solutions include the use of auxiliary pump-storage resonators \cite{hu2022high, buscaino2020design} or soliton crystals \cite{cole2017soliton} to boost the efficiency, at the cost of added system complexity.  Dark solitons also enjoy higher conversion efficiency \cite{xue2015mode, kim2019turn}, but at the cost of narrower bandwidth.  OPO-based combs enjoy $O(1)$ efficiency, derived from the quantum efficiency of the CW OPO and the FMCW pulse shape (Sec.~\ref{sec:2-fmcw}).
	\item {\it Bandwidth.}  Both Kerr and OPO combs can be very broadband due to the virtual transitions that drive the $\chi^{(3)}$ and $\chi^{(2)}$ nonlinearities, although careful dispersion engineering is required for DKS.  In contrast, RE-EO combs are limited by the $V_\pi L\alpha$ of the modulator, while MLL combs are limited by the laser gain medium.
	\item {\it Wavelength tuning.}  DKS and EO combs are locked to the pump, so a broadly tunable source is required for comb tuning.  The FM-OPO can be tuned by finely adjusting the pump detuning (in the manner of a CW-OPO), but the bandwidth decreases as the comb is tuned away from degeneracy \cite{diddams1999broadband, stokowski2024integrated}.  The QAM-OPO enjoys rapid, RF-tunability of the comb that is independent of the pump wavelength (Sec.~\ref{sec:tuning}). 
	\item {\it Spectral flatness.}  Soliton-based combs have a hyberbolic-secant (sech) spectrum that decays exponentially with frequency.  FM combs, i.e.\ the nonresonant EO-comb and FM-OPO, boast a ``flat'' spectrum, but in reality it is a Bessel function that has significant $O(1)$ ripples from line to line \cite{stokowski2024integrated, torres2014optical}.  The QAM-OPO spectrum does not have these ripples.  However, it is also possible to amplitude-modulate an FM comb to smooth out the ripples in the Bessel spectrum \cite{torres2014optical}.
	\item {\it RF power.} Conventional RE-EO combs must drive the EOM hard enough to compensate for cavity loss.  In the OPO comb, this loss is compensated by OPO gain allowing for weaker drive fields.
	\item {\it Turn-on dynamics.}  The Kerr soliton is stabilized by thermal locking \cite{carmon2004dynamical}, and accessing the soliton state is a chaotic, dynamical process that depends on the laser power, tuning rate, and thermal timescale \cite{herr2014temporal} (similarly, passive MLLs often also require careful turn-on dynamics).  Recent efforts to create robust ``turn-key'' solitons through cavity-laser feedback are promising \cite{shen2020integrated, kim2019turn}.  EO and OPO combs exhibit turn-key operation by default.
	\item {\it Cavity locking.} With the exception of the singly-resonant QAM-OPO, all optically-pumped resonant combs in Table~\ref{tab:7-t1} require locking the detuning between the pump laser and cavity mode.  This locking is not necessary in the QAM-OPO.  As a corollary, in the QAM-OPO, pump phase noise is not transferred to the signal comb, allowing low noise combs to be generated from (less expensive) high-noise pump lasers.
\end{itemize}

In conclusion, the QAM-OPO offers a promising new set of capabilities for $\chi^{(2)}$-based optical frequency combs.  Although the intracavity IM makes it more complex than the standard FM-OPO, the field of $\chi^{(2)}$ nanophotonics is rapidly moving in the direction of high-yield, heterogeneous circuits \cite{zhu2021integrated}, where such devices can be fabricated reliably.  This added complexity comes with many benefits, including efficiency, bandwidth, tunability, flatness, low RF power, turnkey operation, and robustness to pump frequency drift, making it a promising source to tackle next-generation scientific \cite{diddams2010evolving, poli2013optical}, and engineering \cite{pfeifle2014coherent, rizzo2022petabit, coddington2016dual, baumann2014comb, trocha2018ultrafast} applications of microcombs.

\section*{Appendix}

\setcounter{equation}{0}
\setcounter{figure}{0}
\setcounter{table}{0}
\setcounter{section}{0}
\makeatletter
\renewcommand{\thetable}{A\arabic{table}}
\renewcommand{\thefigure}{A\arabic{figure}}
\renewcommand{\thesection}{\Alph{section}}
\renewcommand{\thesubsection}{\Alph{section}.\arabic{subsection}}
\renewcommand{\theequation}{A\arabic{equation}}

\section{Notation and Basic OPO Theory} \label{sec:not}

This Appendix contains a derivation of the RE-EO comb spectrum and bandwidth-efficiency tradeoff (App.~\ref{sec:a-eo}), the origin of OPA field equations Eqs.~(\ref{eq:2-d}) (App.~\ref{sec:a-opa}), and calculation of the gain, threshold, bandwidth, and efficiency of a high-finesse, singly-resonant OPO (App.~\ref{sec:a-opo}).

\subsection{Resonant EO Comb} \label{sec:a-eo}

In the absence of dispersion, the EO comb is easy to model in the time domain: let $\alpha$ be the out-coupling efficiency (ignoring other losses) and $\phi \cos(\Omega t)$ be the round-trip phase.  With a CW pump $a_{\rm in}(t) \rightarrow a_{\rm in}$, the steady-state output field is:
\beq
	a_{\rm out}(t) = \Bigl[1 - \frac{\alpha}{1 - \sqrt{1-\alpha}\, e^{i\phi \cos(\Omega t)}}\Bigr] a_{\rm in} \label{eq:a-eoct}
\eeq
We compute the comb spectrum by taking Fourier series integrals of Eq.~(\ref{eq:a-eoct}).  The result, calculated using the contour integrals and the residue theorem, is:
\beq
	a_{m,\rm out} = \biggl\{\delta_{m,0} - \frac{i^{|m|}\,\alpha \exp\Bigl[|m| \sinh^{-1}\Bigl(\frac{\log(1-\alpha)}{2\phi}\Bigr)\Bigr]}{\sqrt{\phi^2 + \bigl(\tfrac12 \log(1-\alpha)\bigr)^2}}
		 \biggr\} a_{\rm in} \label{eq:a-ameo1}
\eeq
For efficient comb generation, the cavity finesse must be large, so the round-trip loss $\alpha$ is small.  Expanding the relevant terms with $\alpha \ll 1$, we get:
\beq
	a_{m,\rm out} = \biggl\{\delta_{m,0} - \frac{i^{|m|}\,\alpha \exp\bigl[-|m| \sinh^{-1}(\alpha/2\phi)\bigr]}{\sqrt{\phi^2 + (\alpha/2)^2}}
		 \biggr\} a_{\rm in} \label{eq:a-ameo2}
\eeq
This expression makes clear that the number of comb lines is related to $\alpha/\phi$.  Large combs require $\phi \gg \alpha$.  Taking this limit gives our final expression for the comb spectrum:
\beq
	a_{m,\rm out} = \Bigl(\delta_{m,0} - \frac{i^{|m|} \alpha}{\phi} e^{-(\alpha/2\phi)|m|} \Bigr)  a_{\rm in} \label{eq:a-ameo3}
\eeq
The 3-dB bandwidth $N_{\text{3dB}}$ (defined in terms of number of comb lines) and comb efficiency $\eta_{\rm comb}$ (defined as comb power divided by input power) are:
\begin{subequations}\begin{align}
	N_{\text{3dB}} & = \frac{f_{\text{3dB}}}{\Omega/2\pi} = 2\log(2)\, \phi/\alpha \label{eq:a-bweo} \\
	\eta_{\rm comb} & = 2\alpha/\phi \label{eq:a-etaeo}
\end{align}\end{subequations}
which is reported at the beginning of the paper in Eq.~(\ref{eq:2-neta}).

Therefore, standard EOM combs suffer from a sad bandwidth-efficiency tradeoff: $N_{\text{3dB}} \eta_{\rm comb} = 4\log(2)$.  This tradeoff is common among ``bright combs'' with pulsed waveforms (although $|a_{\rm out}(t)| = \text{const}$, the non-CW part is a short pulse) in media with time-local nonlinearities, which includes Kerr combs as well \cite{bao2014nonlinear}.  However, by engineering an avoided mode crossing (either to higher-order modes in the ring or with an auxiliary cavity in Vernier configuration \cite{buscaino2020design}), one can recycle the pump and break this tradeoff.

Waveguide loss and modulator half-wave voltage set a fundamental limit to the bandwidth.  A phase modulator's voltage response is given by $\phi = (\pi/2) V/V_\pi$ (the factor of two is a custom arising from the push-pull definition $V_\pi$), so $\phi_p = \phi_{\rm pp}/2 = (\pi/4) V_{\rm pp}/V_\pi$.  The loss is proportional to waveguide length as $\alpha = \alpha_{\rm wg} L$.  Making these substitutions, we find:
\beq
	N_{\text{3dB}} = \frac{\pi \log(2)}{2} \frac{V_{\rm pp}}{V_\pi L\alpha_{\rm wg}} = \underbrace{5\pi \log_{10}(2)}_{\approx\ 5} \frac{V_{\rm pp}\,\text{[V]}}{V_\pi L\alpha_{\rm wg}\, \text{[V\,dB]}} \label{eq:a-bweo2}
\eeq

\subsection{Field Equations in OPA} \label{sec:a-opa}

In this paper, we consider only single-transverse-mode dynamics in the OPA element.  This is appropriate in most situations, particularly if a waveguide is used.  In waveguide nonlinear optics, the electric field can be decomposed into power-normalized modes as follows
\begin{subequations}\begin{align}
	E & = \sum_i e_i \bigl(A_i(z) E_i(x, y)e^{i(\beta_i z-\omega t)} + \mbox{c.c.}\bigr) \\
	H & = \sum_i h_i \bigr(A_i(z) H_i(x, y)e^{i(\beta_i z-\omega t)} + \mbox{c.c.}\bigr) 
\end{align}\label{eq:a-da-nlrep}\end{subequations}
where we choose $e_i = \sqrt{Z_0 n_{g,i}/2}$, $h_i = \sqrt{n_{g,i}/2Z_0}$ to enforce the standard normalization $\int{n n_g |E_i|^2 {\rm d}A} = 1$, while ensuring that power goes as $P = \sum_i |A_i|^2$.

In this paper, we will instead express the $A_i$ in terms of {\it flux-normalized} terms $a_i$, i.e. where the photon flux goes as $J_{\rm ph} = \sum_i |a_i|^2$.  The two are related by factors of $\sqrt{\hbar\omega}$:
\beq
	A_i = \sqrt{\hbar\omega_i}\, a_i
\eeq
An OPA induces a parametric interaction between three fields: pump, signal, and idler.  To begin, consider the CW case, where by convention, these are denoted $C$, $A$, and $B$ (resp.\ $c$, $a$, and $b$), with wavelengths $\omega_c$, $\omega_a$, and $\omega_b$.  The field equations are given by:
\begin{align}
\frac{{\rm d} A}{{\rm d} z} & = i \beta_a A + i \frac{\omega_1}{\omega_3} K^* B^* C, & & & 
\frac{{\rm d} a}{{\rm d} z} & = i \beta_a a + i \kappa^* b^* c, \nonumber \\
\frac{{\rm d} B}{{\rm d} z} & = i \beta_b B + i \frac{\omega_2}{\omega_3} K^* A^* C, & \Leftrightarrow & &
\frac{{\rm d} b}{{\rm d} z} & = i \beta_b b +i \kappa^* a^* c, \nonumber \\ 
\frac{{\rm d} C}{{\rm d} z} & = \underbrace{i \beta_c C}_{\rm linear} + \underbrace{\vphantom{\beta_c}i K A B\ \qquad\ }_{\rm NLO} & & & 
\frac{{\rm d} c}{{\rm d} z} & = i \beta_c c +i \kappa a b \label{eq:a-dandopo}
\end{align}
where $\beta_{a,b,c}$ are the propagation constants, and the nonlinear coefficients $K$ and $\kappa$ are respectively given by:
\begin{align}
K & = \frac{\omega_3}{c} \sqrt{\frac{Z_0 n_{g1}n_{g2}n_{g3}}{2}}\,\chi^{(2)} \int_{\rm NL}{E_3^* E_1 E_2 {\rm d} A} \nonumber \\
\kappa & = \sqrt{\hbar\omega_1\omega_2/\omega_3} K \label{eq:a-kkap}
\end{align}
where the integral $\int_{\rm NL}(\ldots) \d A$ is taken over the nonlinear region.  The contraction the vector indices in the integrand $E_{3,i}^* E_{1,j} E_{2,k}$ are contracted with the elements of the $\chi^{(2)}$ tensor.

To handle ultrashort pulses or frequency combs, we promote the fields to time-dependent quantities, i.e.\ $a(z) \rightarrow a(z, \tau)$, giving the familiar propagative field equations (Eqs.~(\ref{eq:2-d})):
\begin{subequations}\begin{align}
	\partial_z a & = i \beta_a(i\partial_\tau) a + i\kappa\, c b^* \label{eq:a-da} \\
	\partial_z b & = i \beta_b(i\partial_\tau) b + i\kappa\, c a^* \label{eq:a-db} \\
	\partial_z c & = \underbrace{i \beta_c(i\partial_\tau) c}_{\text{Dispersion}} + \underbrace{\vphantom{\beta_c(i\partial_\tau)}i\kappa\, a b}_{\chi^{(2)}} \label{eq:a-dc}
\end{align}\label{eq:a-d}\end{subequations}
We can obtain a related set of equations in Fourier space, where the time-dependent fields are a discrete sum of Fourier modes, i.e.\ $a(\tau) = \sum_m a_m e^{-i\omega\tau}$:
\begin{subequations}\begin{align}
	\frac{\d a_m}{\d z} & = i \beta_a(m \Omega) a_m + i\kappa\sum_n{c_{m+n} b_n^*} \label{eq:a-dfa} \\
	\frac{\d b_m}{\d z} & = i \beta_b(m \Omega) b_m + i\kappa\sum_n{c_{m+n} a_n^*} \label{eq:a-dfb} \\
	\frac{\d c_m}{\d z} & = i \beta_c(m \Omega) c_m + i\kappa\sum_n{a_{n} b_{m-n}} \label{eq:a-dfc}
\end{align}\label{eq:a-df}\end{subequations}
Here $\beta_u(s)$ is the wavenumber at $\omega = \omega_u + s$.  For convenience, we choose to work in the co-propagating rotating-wave basis $\beta_u(s) \rightarrow \beta_u(s) - \beta_a(0) - \beta_a'(0)$, in which $\beta_a(s)$ has only quadratic and higher-order terms.

Since $c_0$ is pumped at CW, most of the relevant processes are of the form $c_0 \leftrightarrow a_m + b_{-m}$.  For simplicity, when studying CW OPOs in the following section, we will denote $a \rightarrow a_m$, $b \rightarrow b_{-m}$, $c \rightarrow c_0$, as well as $\beta_a \rightarrow \beta_a(m\Omega)$, $\beta_b \rightarrow \beta_b(-m\Omega)$, $\beta_c \rightarrow \beta_c(0)$.

\subsection{High-Finesse OPO} \label{sec:a-opo}

\begin{figure*}[t!]
\begin{center}
\includegraphics[width=1.00\textwidth]{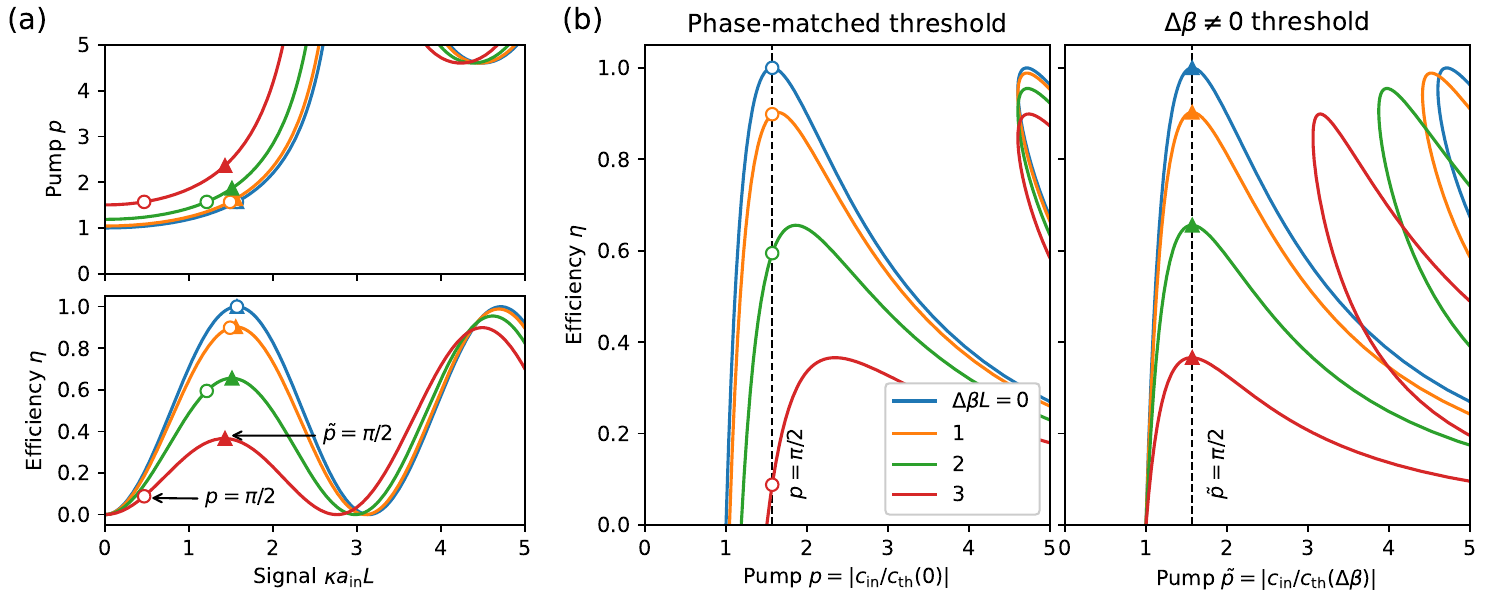} 
\caption{CW efficiency of nondegenerate OPO.  (a) Pump $p$ and efficiency $\eta$ as a function of steady-state power $a_{\rm in}$, Eqs.~(\ref{eq:a-eta}, \ref{eq:a-cth}).  (b) Efficiency vs.\ pump, showing approximate first maximum at $p = \pi/2 \approx 1.57$.}
\label{fig:a-f1}
\end{center}
\end{figure*}

\begin{figure*}[t!]
\begin{center}
\includegraphics[width=1.00\textwidth]{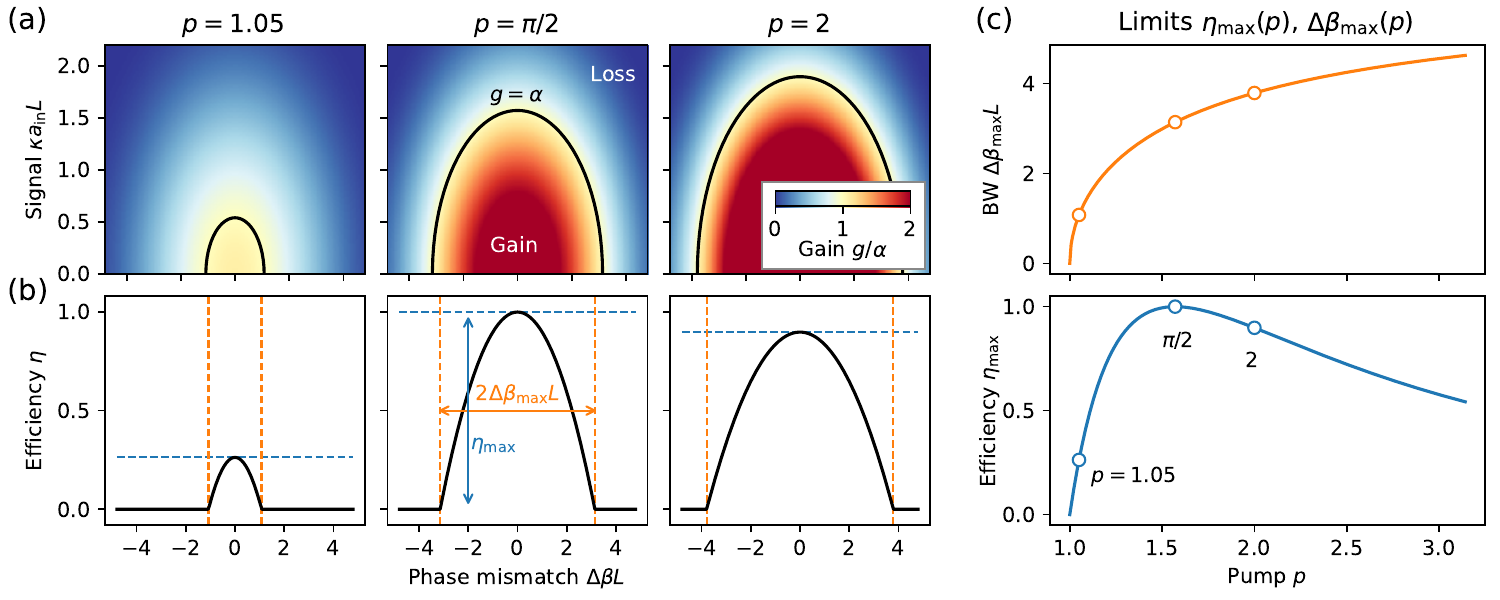} 
\caption{(a) OPA gain as a function of signal amplitude and phase mismatch for three values of $p = 1.05$, $\pi/2$, $2$, Eq.~(\ref{eq:a-gnorm}).  (b) Conversion efficiency $\eta$ at steady-state, Eq.~(\ref{eq:a-etanorm}).  (c) Dependence of $\eta_{\rm max}$ and $\Delta\beta_{\rm max}$ on pump amplitude $p$, Eq.~(\ref{eq:a-parmax}).}
\label{fig:a-f2}
\end{center}
\end{figure*}

For simplicity, here we study the QAM-OPO in the high-finesse limit, which is likely applicable in potential on-chip implementations.  In this limit, $|a| \gg |b|, |c|$, so on a given round trip, the nonlinear effect on $a$ is merely perturbative.  The phase-matched CW solution ($\beta_a = \beta_b = \beta_c = 0$) is:
\begin{subequations}\begin{align}
	a(z) & = a_{\rm in} + \kappa (|c_{\rm in}|^2/2a_{\rm in}) \sin^2(\kappa a_{\rm in} z) \\
	b(z) & = -i c_{\rm in} \sin(\kappa a_{\rm in}z) \\
	c(z) & = c_{\rm in} \cos(\kappa a_{\rm in}z)
\end{align}\end{subequations}
For the phase-mismatched case, the answer is more complicated, and involves applying the substitutions $a = e^{i\beta_a z}\tilde{a}$, $b = e^{i(-\beta_a+\beta_b+\beta_c)z/2} \tilde{b}$, $c = e^{i(\beta_a+\beta_b+\beta_c)z/2} \tilde{c}$, which leads to the following equations:
\beq
	\frac{\d\tilde{a}}{\d z} = i \kappa \tilde{c}\tilde{b}^*,\ \ \ 
	\frac{\d}{\d z} \begin{bmatrix} \tilde{b} \\ \tilde{c} \end{bmatrix}
		= \begin{bmatrix} -\tfrac12i \Delta\beta & i\kappa\tilde{a}^* \\ i\kappa\tilde{a} & \tfrac12i\Delta\beta \end{bmatrix} \begin{bmatrix} \tilde{b} \\ \tilde{c} \end{bmatrix}
\eeq
where $\Delta\beta = \beta_c-\beta_a-\beta_b$ is the phase mismatch.  Defining
\beq
	\gamma = \sqrt{(\Delta\beta/2)^2 + (\kappa a_{\rm in})^2} \label{eq:a-gam}
\eeq
the solution is:
\begin{subequations}\begin{align}
	b(z) & = -i \kappa a_{\rm in} c_{\rm in} \frac{\sin(\gamma z)}{\gamma} e^{i(\beta_b+\Delta\beta)z/2} \\
	c(z) & = c_{\rm in} \Bigl(\cos(\gamma z) + i \frac{\Delta\beta \sin(\gamma z)}{2\gamma}\Bigr)
\end{align}\end{subequations}
The resulting perturbation to $a$ gives:
\begin{align}
	a_{\rm out} & = e^{i\beta_a L} \biggl[1 + \frac{(\kappa|c_{\rm in}| L)^2}{2} \Bigl(\sinc^2(\gamma L) \nonumber \\
	& \qquad\qquad\qquad + i \frac{\Delta\beta}{2\gamma} \frac{1 - \sinc(2\gamma L)}{\gamma L}\Bigr)\biggr] a_{\rm in}
\end{align}
The single-pass gain $g = |a_{\rm out}/a_{\rm in}|^2 - 1 = |b_{\rm out}/a_{\rm in}|^2$ and efficiency $\eta = |b_{\rm out}/c_{\rm in}|^2$ are:
\begin{align}
	g & = (\kappa |c_{\rm in}| L)^2 \sinc^2(\gamma L) \label{eq:a-g} \\
	\eta & = (\kappa |a_{\rm in}| L)^2 \sinc^2(\gamma L) \label{eq:a-eta}
\end{align}
Dividing these expressions, we find the ratio between pump and signal to be:
\beq
	\frac{|a_{\rm in}|}{|c_{\rm in}|} = \sqrt{\eta/g} \label{eq:a-acfrac}
\eeq
This validates the main assumption made at the beginning of Sec.~\ref{sec:a-opo}: in steady-state (when gain equals loss, $\alpha = g$) in the high-finesse limit $\alpha \ll 1$, the signal is much stronger than the pump provided that the efficiency is $O(1)$.

To calculate the OPO pump threshold $c_{\rm th}$, we set $g = \alpha$ in the limit $a_{\rm in} \rightarrow 0$, obtaining:
\beq
	|c_{\rm th}(\Delta\beta)| = \sqrt{\alpha}/(\kappa L\, \sinc(\Delta\beta L/2))
\eeq
This threshold is lowest at phase-matching: $|c_{\rm th}(0)| = \sqrt{\alpha}/\kappa L$.  The pump field relative to threshold is therefore:
\beq
	p \equiv \frac{|c_{\rm in}|}{|c_{\rm th}(0)|} = \frac{1}{\sinc(\gamma L)},\ \ \ 
	\tilde{p} \equiv \frac{|c_{\rm in}|}{|c_{\rm th}(\Delta\beta)|} = \frac{\sinc(\Delta\beta L/2)}{\sinc(\gamma L)} \label{eq:a-cth}
\eeq
Here, the normalized pump can be defined in two ways: relative to the lowest (phase-matched) threshold $p$ and relative to the threshold for the given $\Delta\beta$, $\tilde{p}$.

Note that Eqs.~(\ref{eq:a-eta}, \ref{eq:a-cth}) (plotted in Fig.~\ref{fig:a-f1}(a)) relate the $\eta$ to pump $p$ (or $\tilde{p}$) through $\gamma$, and this relation is independent of $g$.  The form of $\eta(p)$ (and $\eta(\tilde{p})$) is plotted in Fig.~\ref{fig:a-f1}(b).  Usually, one wishes to operate the OPO at maximum efficiency, which for the phase-matched case $\Delta\beta = 0$ corresponds to 100\% at $\tilde{p} = \pi/2$.  As one can see from the figure, the efficiency has multiple peaks, and $\eta \rightarrow 1$ is always possible by taking $\kappa a_{\rm in} L$ (and likewise $p$) large enough.  But this is not realistic in most situations, so considering only the first peak, $\tilde{p} = \pi/2$ is a good approximate (though not exact) optimum for reasonable values of $\Delta\beta \neq 0$.

\begin{figure*}[p]
\begin{center}
\includegraphics[width=1.00\textwidth]{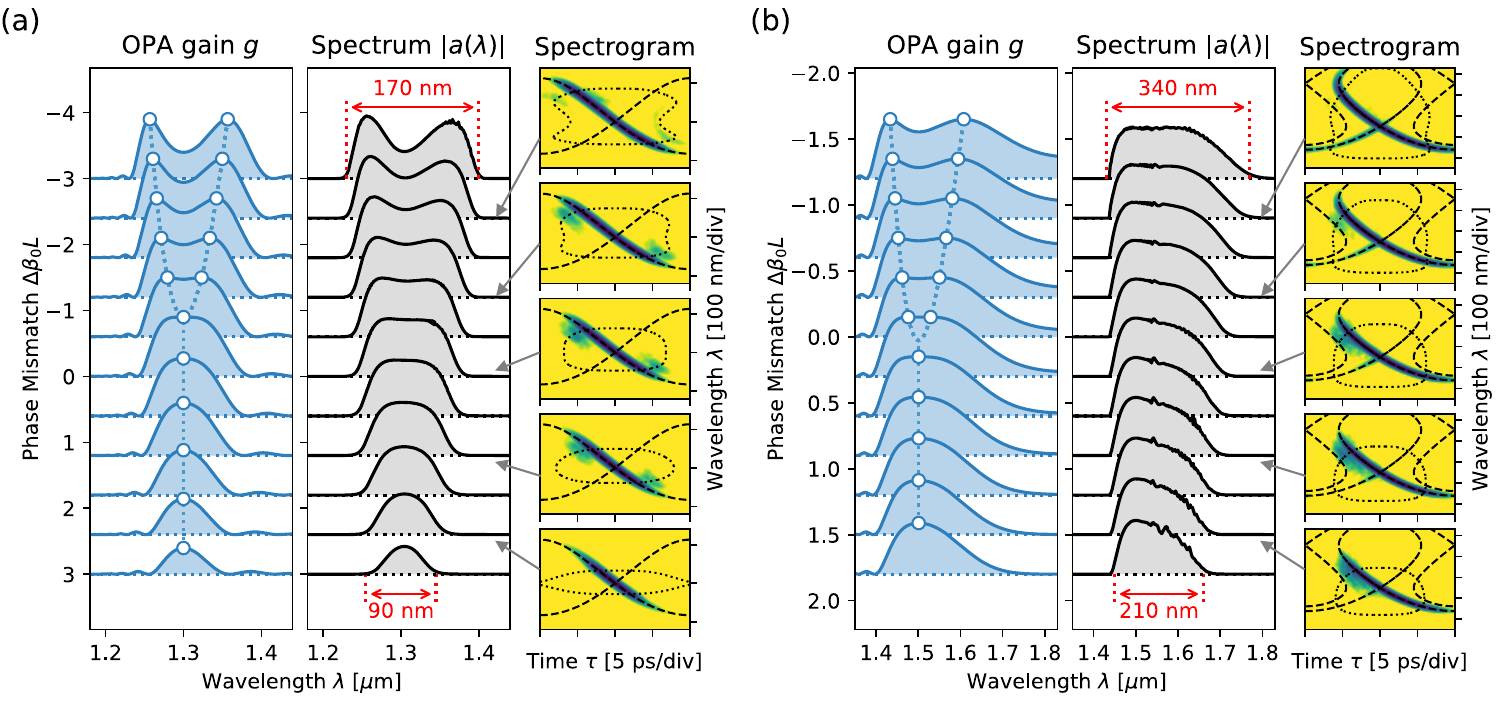} 
\caption{Effect of phase mismatch $\Delta\beta_0 L$ on the OPA gain and signal spectrum for OPOs operating in the O- and C-band.  (a) O-band OPO, $L_{\rm qpm} = 25.7$~$\mu$m, $\lambda_c = 888$~nm, $3\times$ DCF with pure group-delay dispersion.  (b) C-band OPO, $L_{\rm qpm} = 27.4$~$\mu$m, $\lambda_c = 929$~nm, $2\times$ DCF with SMF28 fiber.  Both OPOs use $f_{\rm rep} = 50$~GHz, $\phi_p = 6$~rad.}
\label{fig:b-f1}
\end{center}
\end{figure*}

Recasting the Eq.~(\ref{eq:a-g}) in terms of $p$, we have (Eq.~\ref{eq:2-gnorm})):
\beq
	g = p^2\alpha\, \sinc^2(\gamma L) \label{eq:a-gnorm}
\eeq
In steady-state $g = \alpha$, this yields:
\beq
	\gamma L = \sinc^{-1}(1/p) \sim \begin{cases} \pi/2 & (p = \pi/2) \\ \sqrt{6(p-1)} & (p \approx 1) \end{cases} \label{eq:a-glsinc}
\eeq
Combining Eqs.~(\ref{eq:a-gam}, \ref{eq:a-acfrac}, \ref{eq:a-glsinc}) and thinking very carefully about the math, one derives a formula for the efficiency:
\beq
	\eta = \eta_{\rm max} \bigl[1 - (\Delta\beta/\Delta\beta_{\rm max})^2\bigr] \label{eq:a-etanorm}
\eeq
where
\begin{subequations}\begin{align}
	\eta_{\rm max} & = \Bigl(\frac{\sinc^{-1}(1/p)}{p}\Bigr)^2 \rightarrow
		\begin{cases} 1 & (p = \pi/2) \\ 6(p-1) & (p \approx 1) \end{cases} \\
	\beta_{\rm max} L & = 2\,\sinc^{-1}(1/p) \rightarrow
		\begin{cases} \pi & (p = \pi/2) \\ 2\sqrt{6(p-1)} & (p \approx 1) \end{cases}
\end{align}\label{eq:a-parmax}\end{subequations}
Eqs.~(\ref{eq:a-gnorm}-\ref{eq:a-parmax}) are plotted in Fig.~\ref{fig:a-f2}.  This shows the net gain as a function of both resonant signal and phase mismatch, as well as the relation between phase mismatch and efficiency.  Maximum efficiency is achieved at $p = \pi/2$, and the bandwidth $\Delta\beta_{\rm max}$ increases for larger pump amplitudes, though not significantly.

\section{Bandwidth Optimization} \label{sec:bw}

\begin{figure*}[p]
\begin{center}
\includegraphics[width=1.00\textwidth]{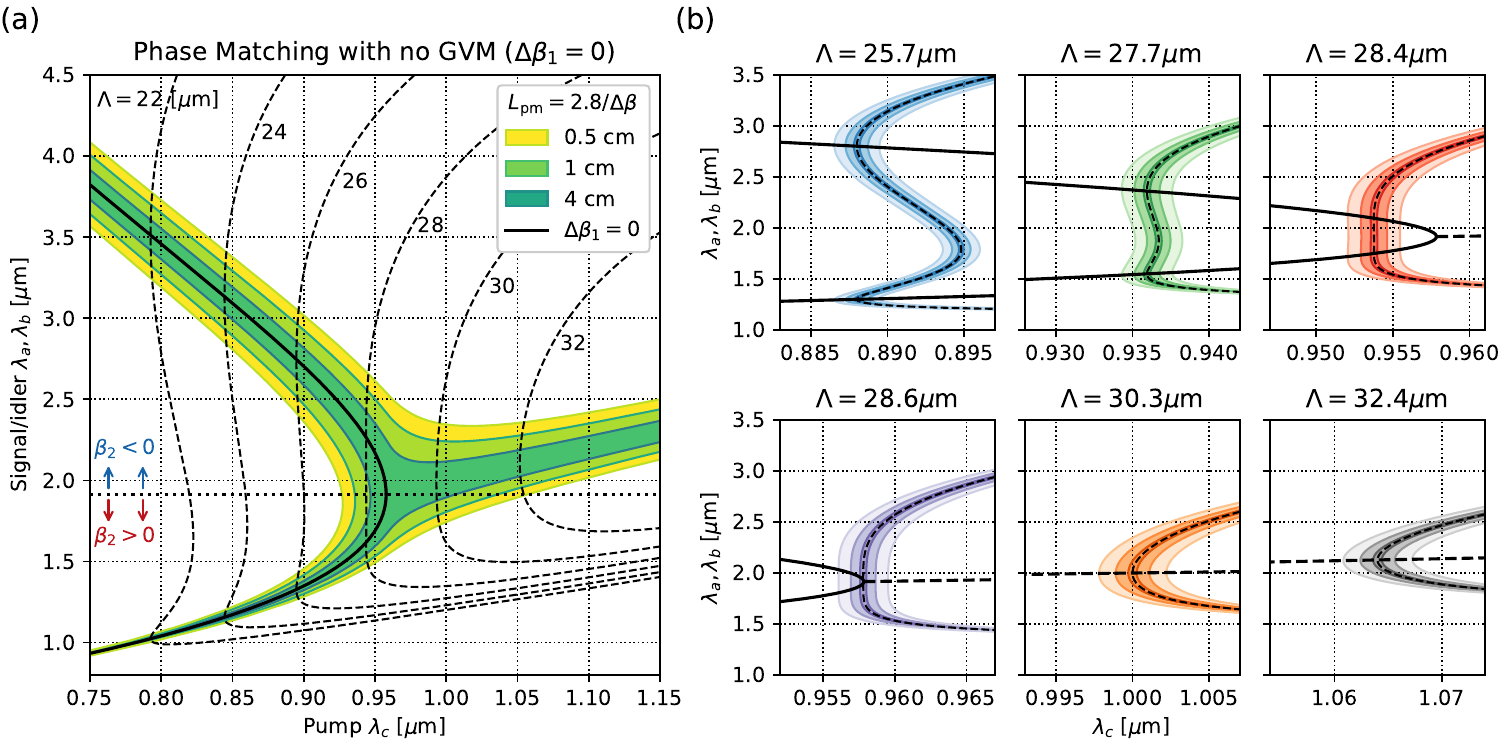} 
\caption{(a) Group-velocity matching condition for bulk LiNbO$_3$, and 3-dB gain spectra for crystals of different sizes, under the condition $\Delta\beta_1 = 0$.  (b) Gain spectra as a function of pump for six different QPM lengths.}
\label{fig:b-f2}
\end{center}
\end{figure*}

Recall that the undepleted gain $p^2\alpha\,\text{sinc}^2(\Delta\beta(\omega) L/2)$ in an OPO (Eq.~(\ref{eq:a-gnorm})) is governed by both the pump power and phase matching, the latter given by:
\beq
	\Delta\beta(\omega) = \beta(\omega_c) - \beta(\omega) - \beta(\omega_c-\omega) - \frac{2\pi}{\Lambda_{\rm QPM}} \label{eq:b-gain}
\eeq
Here $p = c/c_{\rm th}$ is the normalized pump, $\alpha$ is the round-trip loss, $\omega_c$ and $\omega$ are the pump and signal wavelengths, and $\Delta\beta(\omega)$ is the phase-mismatch term for the process $\omega_c \rightarrow (\omega, \omega_c-\omega)$, with gain maximized when $\Delta\beta = 0$.  Let $\omega_a$ be the target center frequency.  Expanding about $\omega_a$, i.e.\ $\omega = \omega_a + \Delta\omega$, we find:
\begin{align}
	& \Delta\beta \nonumber = \Delta\beta(\omega_a) + \underbrace{(\beta_{1,a} - \beta_{1,b})}_{\Delta\beta_1} \Delta\omega \nonumber \\
	& \qquad\qquad + \underbrace{\tfrac{1}{2}(\beta_{2,a} + \beta_{2,b})}_{\bar{\beta}_2} \Delta\omega^2 + O(\Delta\omega^3) \label{eq:b-db}
\end{align}
The goal of bandwidth optimization is to keep the gain $g(\omega)$ large (equivalently, keep $\Delta\beta(\omega)$ small) over as wide a frequency window as possible.  Examining Eqs.~(\ref{eq:b-gain}-\ref{eq:b-db}), we arrive at the following operation guidelines for the OPO:
\begin{itemize}
	\item The target frequency $\omega_a$ should either be a zero or a minimum of $\Delta\beta(\omega)$, in order for the gain to be maximized at $\omega_a$.
	\item The signal-idler GV mismatch $\Delta\beta_1 = \beta_{1,a} - \beta_{1,b}$ should be minimized or set to zero in order to maximize the phase-matching bandwidth.
	\item In addition, the mean signal-idler GVD $\bar{\beta}_2 = \tfrac12 (\beta_{2,a} + \beta_{2,b})$ should be minimized if this is possible.  However, by Taylor expanding $\beta(\omega)$ about the center frequency $\bar{\omega} = \tfrac12(\omega_a + \omega_b)$, we find:
	\beq
		\bar{\beta}_2 = \frac{\Delta\beta_1}{\omega_a-\omega_b} + \frac{1}{12} \beta_4(\bar{\omega}) (\omega_a-\omega_b)^2 + O\bigl((\omega_a-\omega_b)^4\bigr)
	\eeq
	Plugging this into Eq.~(\ref{eq:b-db}), we find the following approximation for $\Delta\beta$:
	\begin{align}
		\Delta\beta & = \Delta\beta(\omega_a) + \Delta\beta_1 \Delta\omega \nonumber \\
		& \quad + \Bigl[\frac{\Delta\beta_1}{\omega_a - \omega_b} + \frac{\beta_4(\bar{\omega})}{12} (\omega_a-\omega_b)^2 \Bigr] \Delta\omega^2 + \ldots \label{eq:b-db2}
	\end{align}
\end{itemize}
The 3-dB gain bandwidth is obtained by solving Eq.~(\ref{eq:b-db2}) for $\Delta\beta(\Delta\omega_{\rm max}) = \Delta\beta_{\rm max}$, where $\Delta\beta_{\rm max} = (2/L)\text{sinc}^{-1}(1/\sqrt{2}) \approx 2.8/L$ per Eq.~(\ref{eq:b-gain}).  This is limited by the leading order in the Taylor expansion Eq.~(\ref{eq:b-db2}), i.e.\ $\Delta\beta_{1}$ in the case of group-velocity mismatch, and $\beta_4(\bar{\omega})$ in the case that the group velocities match.  Moreover, when the group velocities match, it is because the signal and idler are approximately on opposite sides of the zero-(group)-dispersion wavelength, so we have $\bar{\omega} \approx \omega_{\rm zdw}$.  Assuming perfect phase-matching at $\omega_a$, the bandwidth limits are as follows:
\beq
	\text{BW}_{\rm gain} = \begin{dcases}
		\frac{2\,\Delta\beta_{\rm max}}{|\Delta\beta_1|} & \Delta\beta_1 \neq 0 \\
		\frac{4\sqrt{3\Delta\beta_{\rm max}/|\beta_4(\omega_{\rm zdw})|}}{\omega_a - \omega_b} & \Delta\beta_1 = 0
		\end{dcases} \label{eq:b-bw}
\eeq
Re-expressing in terms of wavelengths, this is:
\beq
	\Delta\lambda_{\rm gain} 
		= \begin{dcases}
			\frac{\lambda_a}{\pi c} \frac{\Delta\beta_{\rm max}}{|\Delta\beta_1|} & \Delta\beta_1 \neq 0
			 \\
			\frac{\lambda_a^3\lambda_b}{\lambda_b-\lambda_a} \frac{\sqrt{3\Delta\beta_{\rm max}/|\beta_4(\omega_{\rm zdw})|}}{(\pi c)^2} & \Delta\beta_1 = 0
		\end{dcases} \label{eq:b-bwl}
\eeq
For bulk LiNbO$_3$ with $\lambda_{\rm zdw} = 1.92$~$\mu$m, we find that an OPO with group-velocity matching will have a bandwidth of $\Delta\lambda_{\rm gain} = 100$~nm and 270~nm for signals at the O-band (1.3~$\mu$m) and C-band (1.5~$\mu$m), respectively.

We can actually do a little better than this by slightly phase-mismatching the nonlinear interaction, in order to create a double-peaked gain spectrum as shown in Fig.~\ref{fig:b-f1}.  In particular, if we set $\Delta\beta(\omega_a) = -\text{sign}(\beta_4) \Delta\beta_{\rm max}$, then we can expand the gain window by a factor of $\sqrt{2}$, i.e.\ to 140~nm and 380~nm for the O- and C-band OPOs, respectively.  We see this agrees pretty well for the O-band OPO (Fig.~\ref{fig:b-f1}(a)), although it comes at the cost of a double-peaked spectrum with a dip at 1.3~$\mu$m (spectra are shown on a linear scale).  Further increasing the phase mismatch causes the two gain regions to completely decouple, with the saddle point falling out of the gain region.  The OPO becomes chaotic in this regime, since there is no optical power at the saddle point to connect the two gain lobes; instead, the two gain regions independently amplify vacuum noise without any stable phase relation.

A similar picture plays out for the C-band OPO, where a slight phase mismatch increases the comb bandwidth.  In this case, however, with a 340~nm comb, power already extends all the way to the degeneracy point (about 1.85~$\mu$m).  Moreover, loopback instability also comes into play for such broad combs, so higher-order dispersion compensation will be critical to exploiting this broad gain bandwidth if a reasonable modulation phase is to be used.

Fig.~\ref{fig:b-f2} helps guide our thought process for picking conditions that maximize OPO bandwidth.  Given a particular gain material (using bulk LiNbO$_3$ as an example), we can find the signal-idler group-velocity matching condition as a function of pump and signal (solid curve in Fig.~\ref{fig:b-f2}(a)).  In order to match these, the signal and idler must be on opposite sides of the zero-dispersion wavelength $\lambda_{\rm zdw}$, and this constrains GVM-free operation to OPOs pumped below $0.96$~$\mu$m.  We also plot the phase-matching bandwidth, assuming group-velocity matching, confirming the trend of Eqs.~(\ref{eq:b-bw}-\ref{eq:b-bwl}) that the bandwidth is broader when the signal and idler wavelengths are closer.

Of course, we are not restricted to using the crystal at the group-velocity matched condition.  Fig.~\ref{fig:b-f2}(b) shows the gain conditions for a set of different crystal QPM periods, where we are free to vary the pump wavelength.  Operating on the $\Delta\beta_1 = 0$ line gives the maximum bandwidth, but in many cases, we can still get moderately broadband gain when deviating slightly, at least when pumping to the blue of $\lambda_{\rm zdw}/2$.  Note that, in this case, there are also {\it two} signal-idler pairs that can oscillate.  These pairs both satisfy phase-matching, so theoretically have the same gain and can compete with each other for power.  When forming an OPO comb, usually only one of these gain peaks will contain the separatrix and saddle point, and this is where the comb originates.  The phase-space dynamics, specifically group-delay mismatch, will usually (but not always) lead to suppression of any signals amplified by the competing gain peak.

\section{QAM-OPO in Degenerate Operation} \label{sec:deg}

\begin{figure*}[t]
\begin{center}
\includegraphics[width=1.00\textwidth]{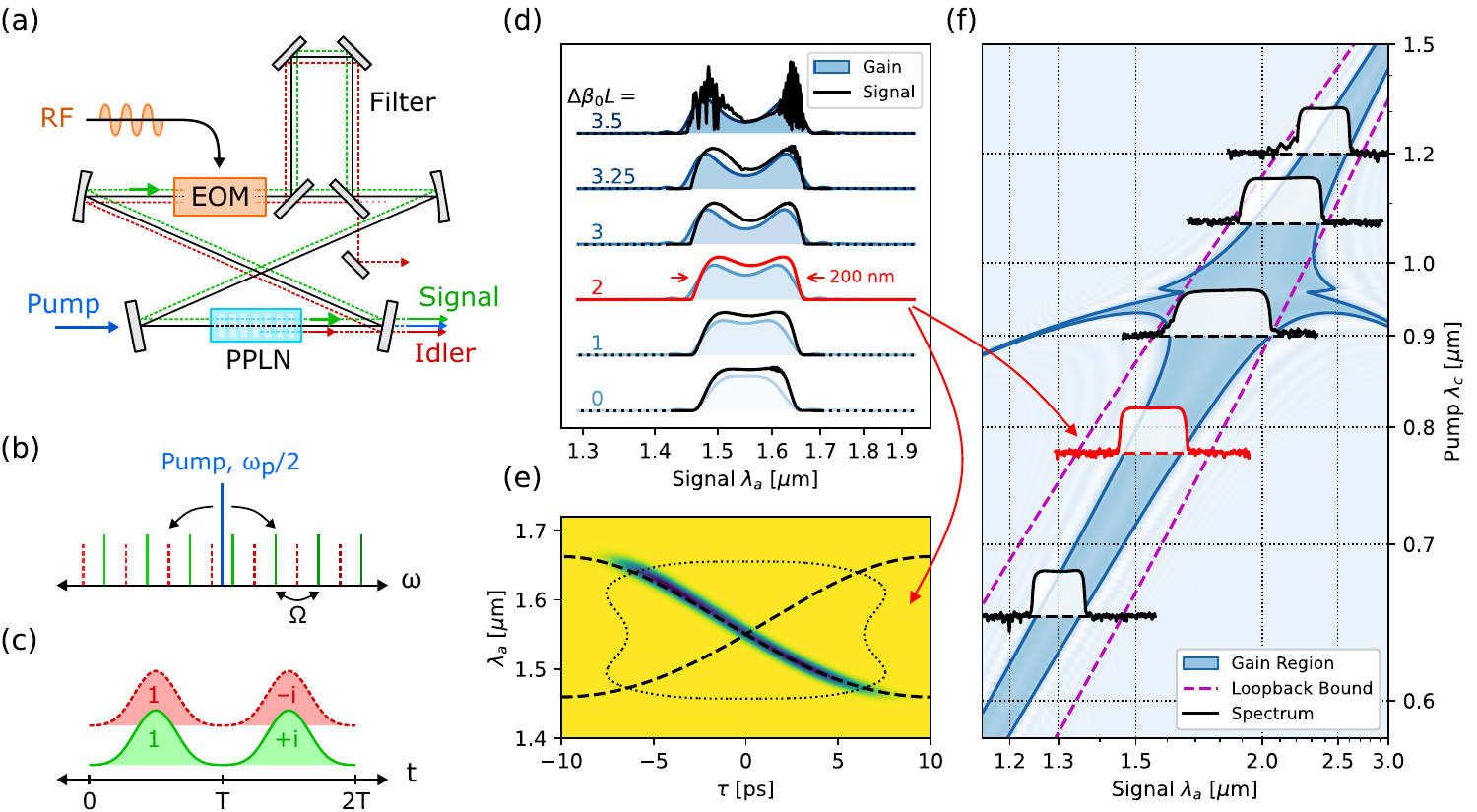} 
\caption{Degenerate operation of an EO Comb OPO.  Effective SROPO at the degeneracy point is enabled using (a) a delay-line filter, and (b) pumping in the $\pi$-phase condition, which is equivalent to (c) the signal and idler bands occupying orthogonal temporal modes.  (d) Simulated comb spectra for a degenerate ECPO with signal centered at 1.55~$\mu$m.  (e) Corresponding spectrogram.  (f) Gain-bandwidth and loopback conditions limit the bandwidth of the generated combs.  Five example spectra centered at $\lambda_a = 1.3~\mu$m, 1.55~$\mu$m, 1.8~$\mu$m, 2.128~$\mu$m, and 2.4~$\mu$m.  ECPO parameters: $f_{\rm rep} = 50$~GHz, $\phi_p = 6.0$, $p = 1.8$, $\alpha = 0.2$.  $\Delta T$, $\Delta\beta_0$, $\phi_a$, and auxiliary GDD tuned to optimize each spectrum individually.}
\label{fig:c-f8}
\end{center}
\end{figure*}

The QAM-OPO requires singly-resonant operation, with a resonant signal and non-resonant idler band.  The most common way to make an OPO singly-resonant is to pump it nondegenerately, so the signal and idler are at separate frequencies.  A degenerate OPO with type II phase matching (with signal and idler in orthogonal polarizations) can also be used, since in this case one can filter out the idler with a polarizer.  However, for both of these cases, the phase-matching bandwidth is usually relatively narrow unless the system is engineered to match group velocities (e.g.\ Fig.~\ref{fig:b-f2}(a)), which often leads to inconvenient pump and idler wavelengths.

Another strategy enforce singly-resonant operation at degeneracy, shown in Fig.~\ref{fig:c-f8}(a), is to add a periodic filter (with period $2\times \text{FSR}$) in the cavity.  This filter (illustrated as a delay line, but an etalon would also work) splits the spectrum into two interleaved mode pairs (signal and idler), transmitting one and dumping the other.  We pump the OPO in the $\pi$-phase condition (Fig.~\ref{fig:c-f8}(b)), which means that the pump half-harmonic $\omega_p/2$ is halfway between two adjacent cavity resonances.  This causes the degenerate OPA crystal to effectively implement a nondegenerate interaction between the signal and idler modes, which in the time domain have the phase relation traced in Fig.~\ref{fig:c-f8}(c) between adjacent round trips.  The nonlinear dynamics are the same as for a nondegenerate QAM-OPO---the only difference is that we now have $\omega_a = \omega_b$ in the phase-matching condition.

Degeneracy is a very effective way to for broaden the phase-matching bandwidth (a property exploited by FM-OPO combs \cite{diddams1999broadband, stokowski2024integrated}), since it zeroes out the group-velocity mismatch $\Delta\beta_1$ (Eq.~(\ref{eq:b-db})), removing the leading term in the phase mismatch (as well as any higher-order odd terms):
\beq
	\Delta\beta = \Delta\beta_0 + \beta_{2,a} \Delta\omega^2 + \frac{\beta_{4,a}}{12} \Delta\omega^4 + O(\Delta\omega^6) \label{eq:c-dbdeg}
\eeq
Thus, the phase-matching bandwidth is $\Delta\omega_{\rm BW} = 2\sqrt{\Delta\beta_{\rm max}/\beta_{2,a}}$ in the phase-matched case ($\Delta\beta_0 = 0$), and by introducing a small initial phase mismatch $\Delta\beta_0 = -\text{sign}(\beta_{2,a}) \Delta\beta_{\rm max}$ (see Appendix~\ref{sec:bw} above), we can increase this by a factor of $\sqrt{2}$:
\beq
	\Delta\omega_{\rm BW} \leq \begin{cases}
		2\sqrt{\Delta\beta_{\rm max}/\beta_{2,a}} & (\Delta\beta_0 = 0) \\
		2\sqrt{2\Delta\beta_{\rm max}/\beta_{2,a}} & (\Delta\beta_0 = -\text{sign}(\beta_{2,a}) \Delta\beta_{\rm max}) \end{cases} \label{eq:c-dwdeg}
\eeq
Further increasing $\Delta\beta_0$ leads to nondegenerate operation.

Fig.~\ref{fig:c-f8}(d) shows representative QAM-OPO spectra when operated at degeneracy (with signal $\lambda_a = 1.55~\mu$m) with various values of $\Delta\beta_0$, showing the expected broadening, followed by a transition to nondegeneracy and unstable operation.  The flat-top comb has a bandwidth of 200~nm, in agreement with the predicted value from Eq.~(\ref{eq:c-dwdeg}).  This comb shows a typical FMCW spectrogram (Fig.~\ref{fig:c-f8}(e)), where the separatrices trace out clean sinusoids showing that the inverted pendulum model is valid here.

What about other wavelengths?  In Fig.~\ref{fig:c-f8}(f), we plot the gain region (defined as the 3-dB window of the gain spectrum, i.e.\ where $g(\omega) > \tfrac12 g_{\rm max}$) as a function of pump and signal wavelength, assuming degeneracy with a slight phase mismatch $\Delta\beta_0 L = -\text{sign}(\beta_{2,a})$ to improve the bandwidth.  The bandwidth is broadest near the zero-dispersion wavelength, where Eq.~(\ref{eq:c-dwdeg}) formally diverges.  However, we are unable to exploit the full bandwidth in this region due to loopback instability, which limits the bandwidth to (Eq.~(\ref{eq:3-lpmax})):
\beq
	\Delta\omega_{\rm BW} \leq 3\Bigl(\frac{3\phi_p}{\delta_3}\Bigr)^{1/3}
\eeq
assuming tunable GVD while the TOD of the system is fixed.  (Note that compensating both GVD and TOD will lead to a larger bound, eventually limited by $4^{\rm th}$ and higher-order dispersion terms, depending on how much dispersion engineering we want to do).  To show that we can saturate this gain window for realistic systems, in Fig.~\ref{fig:c-f8}(f), we overlay degenerate QAM-OPO signal spectra for five representative wavelengths: $1.3~\mu$m, $1.55~\mu$m, $1.8~\mu$m, $2.128 = 2\times 1.064~\mu$m, and $2.4\mu$m (on a log-scale, so the signals look flatter than in Fig.~\ref{fig:c-f8}(d)).

Finally, the {\it periodicity of the filter} will limit the useful bandwidth of this degenerate QAM-OPO, since we require a filter whose transmission maxima and minima are evenly spaced over the full comb bandwidth.  A delay-line filter like that in Fig.~\ref{fig:c-f8} will have a transmission spectrum that goes as
\beq
	T(\omega) = \frac{1 + e^{i\beta(\omega)L}}{2} = e^{i\beta(\omega)L/2} \cos\bigl(\beta(\omega)L/2\bigr)
\eeq
where $L$ is the length of the delay line and $\beta(\omega)$ is the dispersion relation of the waveguide.  The transfer function amplitude will be perfectly periodic for a linear dispersion relation, with a period $\Omega = 2\pi/\beta_1 L = 2\pi c/n_g L$.  This periodicity is broken by GVD and higher-order dispersion terms, which impart an additional delay-line phase $\Delta\phi = (\tfrac12 \beta_2 \Delta\omega^2 + \tfrac16 \beta_3 \Delta\omega^3 + \ldots)L$.  This phase shift affects the filter's behavior away from the center of the comb: a fraction $\sin^2(\Delta\phi/2)$ of the idler is recycled through the cavity (rather than being dumped on each round trip), and the signal is attenuated by a factor of $\cos^2(\Delta\phi/2)$ per round-trip.  Given a maximum phase-error tolerance $\Delta\phi_{\rm max}$, the comb bandwidth is bounded by:
\beq
	\Delta\omega_{\rm BW} \leq \begin{cases}
		2\sqrt{\beta_1 \Omega\,\Delta\phi_{\rm max}/\pi \beta_2} & (\beta_2 \neq 0) \\
		2(3\beta_1 \Omega\,\Delta\phi_{\rm max}/\pi \beta_3)^{1/3} & (\beta_2 = 0)
	\end{cases} \label{eq:c-dwfilter}
\eeq
For example, for the 1.55~$\mu$m QAM-OPO studied in Fig.~\ref{fig:c-f8}(d), assuming a delay line with the same dispersive properties as SMF28 fiber ($n_g = 1.47$, $\beta_2 = -20$~fs$^2$/mm) and a maximum phase error of $\pi/2$, Eq.~(\ref{eq:c-dwfilter}) imposes a bandwidth limit of $\Delta\omega_{\rm BW} \leq 60$~THz, about 500~nm.  The generated QAM-OPO spectrum is well within this limit.  Note, however, that using a more dispersive waveguide for the filter could significantly reduce $\Delta\omega_{\rm BW}$, so the delay-line filter is one of the components that must be carefully dispersion engineered for a broadband system.

While degenerate operation does not yield spectra that are intrinsically more broadband than those studied previously in Appendix~\ref{sec:bw}, it offers several practical advantages:
\begin{itemize}
	\item It operates with a different (and perhaps more convenient) pump wavelength.  Broadband nondegenerate phase-matching relies on group-velocity matching, ideally with a signal and idler close to the zero-dispersion wavelength.   We find in practice (see Fig.~\ref{fig:b-f2}) that these conditions constrain the pump to a really inconvenient wavelength range $\lambda_c \in [0.85, 0.95]~\mu$m.  If one does not wish to pump at this wavelength for some reason, it may be better to operate at degeneracy.
	\item No separate idler band.  As a result, the devices only need to be designed with two frequency bands in mind, not three.  Note also that the group-velocity matching condition in Fig.~\ref{fig:b-f2}(a) requires very long idler wavelengths if the signal wavelength is short; in LiNbO$_3$, this can lead to absorption in the core or the oxide cladding.
	\item You get double the number of comb lines.  Recall from the mode-hop stability analysis (Sec.~\ref{sec:3-modehop}) that the number of comb lines is limited by mode-hopping effects, and scales as $N_{\rm comb} \propto \phi_p/\alpha$.  This analysis only counts {\it signal} comb lines, but here the signal and idler combs are interleaved, so the total number of lines is doubled.
	\item Higher conversion efficiency, since both the signal and idler combs are at the same desired wavelength (these combs can be re-interleaved with a delay-line lattice filter like that used in Fig.~\ref{fig:c-f8}(a)).
\end{itemize}

%


%
\bibliography{refs}{}
\bibliographystyle{IEEEtranN}

\end{document}